\documentclass[preprint,review,12pt]{article}

\usepackage{amsmath, amssymb}
\usepackage[latin1]{inputenc}
\usepackage{enumitem}
\usepackage{graphicx}
\usepackage{caption}
\usepackage{textcomp}
\usepackage{subcaption}

\usepackage{cancel}
\usepackage{authblk}
\usepackage{amsmath}
\usepackage[numbib]{tocbibind}
\usepackage[colorlinks=true,linkcolor=black, citecolor=blue, urlcolor=blue]{hyperref}

\begin{document}
%\begin{frontmatter}

\title{Gradient models of moving heat sources for powder bed fusion applications}
\author[1,2]{Y.O. Solyaev}
\author[1,2]{S.A. Lurie}

%\author{Yury Solyaev\corref{cor1}}
%\cortext[cor1]{Corresponding author}
%\ead{yos@iam.ras.ru}
%\author[label1,label2]{Sergey Lurie}
%\author[label3]{Antipov V.}
%\author[label3]{Serebrennikova N.}
%\author[label2]{Prokudin O.}

\affil[1]{Institute of Applied Mechanics of Russian Academy of Sciences, Moscow, Russia}
\affil[2]{Moscow Aviation Institute, Moscow, Russia}

\setcounter{Maxaffil}{0}
\renewcommand\Affilfont{\itshape\small}
% Collaboration name, if desired (requires use of superscriptaddress option in \documentclass). 
% \noaffiliation is required (may also be used with the \author command).
%\collaboration{}
%\noaffiliation

\date{\today}
\maketitle
%\address[label3]{All-Russian Institute of Aviation Materials, Moscow, Russia}

\begin{abstract}
In this paper, we derive closed form solutions for the quasi-stationary problems of moving heat sources within the gradient theory of heat transfer. This theory can be formally deduced from the two-temperature model and it can be treated as a generalized variant of the Guyer-Krumhansl model with the fourth order governing equation. We show that considered variant of the gradient theory allows to obtain a useful singularity-free solutions for the moving point and line heat sources that can be used for the refined analysis of the melt pool shape in the laser powder bed fusion processes. Derived solutions contain single additional length scale parameter that can be related to the mean particles size of the powder bed.  Namely, we show that developed gradient models allow to describe the decrease of the melt pool depth with the increase of the powder's particles size that was observed previously in the experiments. We also derive the dimensionless relations that can be used for the experimental identification of the model's length scale parameter for different materials. Semi-analytical solution for the Gaussian laser beam is also derived  and studied based on the Green function method within the considered theory.
\end{abstract}

%%%%%%%%%%%%%%%%%%%%%
\section{Introduction}
\label{intro}
Analytical and numerical models of moving heat sources are widely used in the simulations of 
%different technological processes, such that 
welding \cite{arora2019thermal}, laser treatment and cutting \cite{gladush2011physics},  frictional systems \cite{laraqi2003exact}, laser and electron beam additive manufacturing \cite{gusarov2009model,promoppatum2017comprehensive}. 
Pioneering works for the quasi-stationary problems with moving heat sources  have been performed theoretically by Wilson \cite{wilson1904} and later by Rosenthal with application to welding \cite{rosenthal1946theory}.
More complex transient solutions and models with different shapes of the heat sources  and the heat intensity distributions have been later developed \cite{eagar1983temperature,hou2000general}. 

Today, the moving heat source models are the basis for the simulations of the powder bed fusion processes. Such models allow to evaluate the thermal state of the parts during 3d-printing and subsequently to predict the related effects in the change of materials microstructure and porosity, phase composition, surface roughness, over-melting effects and the residual stress state \cite{yan2018review,srivastava2020multi,solyaev2019overmelting}. The most common approach is to use the Gaussian-type models for the laser beam together with numerical simulations of the conjugate heat and mass transfer inside the melt pool and in the heat affected zone \cite{yan2018review}. Simplified quasi-stationary and point-source analytical and semi-analytical models are also used to obtain approximate predictions for the melt pool morphology and to select  the optimal values of the process parameters (laser power, scanning speed, spot size, etc.) \cite{promoppatum2017comprehensive,tian2020melt}. Analytical solutions are also involved in the novel hybrid methods for the improvement of computational efficiency of the long-term finite-element simulations \cite{moran2021scan}. 

In the present paper we developed a new class of the moving heat source models that allow to take into account the effect of the material microstructure. 
Obtained analytical and semi-analytical solutions are developed within the high-order heat transfer theory. This theory can be formally deduced from the 
known mixture theory of the heat conducting two-component systems 
(or the two-temperature model) \cite{forest2010some}. Also, considered theory can be obtained based on the extended approaches of continuum thermodynamics, 
assuming that the free energy density depends not only on the temperature field and its first gradients, but also on its second gradients \cite{forest2010some,nguyen2005non} (that's why we called the derived solutions as the "gradient models"). The governing equation of the considered model, i.e. the heat equation, has the fourth order with respect to the spatial coordinates and the second order with respect to time. This model is the generalized variant of Cattaneo's and Guyer-Krumhansl's models (see Ref.\cite{forest2010some}). 

In order to derive closed form analytical solutions we introduce  simplified constitutive assumptions about the relations between the gradient parameters of the model. As the result, the developed solutions contain single additional length scale parameter that can be treated as the material's internal characteristic length scale. Similar parameters always arise in the high-order gradient theories that are well established today in elasticity \cite{dell2009generalized}, in hydrodynamics \cite{fried2006tractions,rosi2013propagation}, in electrodynamics \cite{lazar2020second}, and in different coupled field theories \cite{sciarra2007second,solyaev2021electric}. Second gradient modification of the Newtonian gravity theory have been also discussed recently in Ref. \cite{lazar2020gradient}. 

Identification of the length scale parameters of gradient theories is a specific problem that have been solved previously, for example, within the gradient elasticity theory for the single crystals and idealized structures based on the molecular dynamics\cite{lurie2017identification} and first principle calculations \cite{shodja2013ab}. Discrete and numerical models have been used to evaluate these parameters for the inhomogeneous materials and metamaterials \cite{yvonnet2020computational,abdoul2018strain,dell2016large}. 
Experimentally observed size effects have been also used for the parameters identification of  gradient models of nano-composites and nano-fluids \cite{ma2018inclusion,solyaev2020gen}. 

Generally, the need for the use of high-order gradient continuum theories can be justified for the precesses with relatively high gradients of the field variables. Such situations arise when one consider the small scale systems \cite{cordero2016second}, microstructured materials and metamaterials \cite{eremeyev2018linear}, and the high frequency phenomena \cite{askes2011gradient}. Giant spatial and temporal gradients of temperature in the laser treatment processes make them a promising area of application of gradient theories. For simulations of such processes the two-temperature model and the Guyer-Krumhansl's model have been widely established \cite{kovacs2018analytic,naldo2020understanding}. In the present paper, we show that phenomenological gradient models of heat transfer can be also useful for the refined analysis of the melt pool morphology that forms during the laser powder bed fusion.

The rest part of the paper is organized as follows. In Section 2 we consider the  gradient theory of heat transfer and propose its particular variant, which allows an explicit definition of general solution for moving source problems. In Section 3 we derive a closed form solutions for the point, line and Gaussian heat sources. Dimensionless forms of these solutions are given. Method for identification of the additional length scale parameter of the model is proposed. In Section 4 we present the results of numerical calculations for the temperature profiles and the melt pool shape based on the obtained solutions for moving source problems. 
Comparison of the model with available experimental data and example of identification of the model's additional length scale parameter for the tungsten powder are presented. 

\section{Gradient model of heat transfer}
\label{sec1}
\subsection{Governing equation}
\label{sec1a}

Consider the heat conduction in the infinite isotropic and homogeneous medium. High-order heat equation within the considered gradient theory of heat transfer can be expressed in the following form \cite{forest2010some}:
\begin{equation}
\label{HE}
(1 - g^2 \Delta)\dot \theta + \tau \ddot \theta = \kappa (1 - l^2 \Delta) \Delta \theta + \frac{1}{c\rho}\hat q
\end{equation}
where $\theta(\textbf x) = T - T_i$ $[K]$ is the rise of temperature $T(\textbf x)$ over the initial level $T_i$ at a point $\textbf x = \{x,y,z\}$; $\kappa$ $[m^2/s]$ is thermal diffusivity, $c$  $[J/(kg \,K)]$ is the heat capacity at constant pressure, $\rho$ $[kg/m^3]$ is the mass density, $\tau$ $[s]$ is relaxation time 
$g$ $[m]$ is the so-called dissipation parameter \cite{fulop2018emergence},
$l$ $[m]$ is the additional length scale parameter of gradient theory, $\hat q$ $[J/m^3]$ is the volumetric power source; $\Delta$ is three-dimensional Laplacian operator and $\Delta\Delta = \Delta^2$ is the biharmonic operator.

In comparison with standard Guyer-Krumhansl model \cite{fulop2018emergence}, equation \eqref{HE} contains additional biharmonic term $\Delta^2\theta$. 
Previously, it was shown\cite{forest2010some} that equation \eqref{HE} straightforwardly follows from the 
theory of the two-component heat conducting mixtures (or 
the two-temperature model) 
if one assumes that the averaged temperature can be approximated by arithmetic mean of the individual temperatures of the components.
We can also note that the governing equation of the form \eqref{HE} is similar to those one that arise in the theory of multi-component diffusion \cite{aifantis1979new}. Biharmonic term $\Delta^2\theta$ also arises in the heat transfer models with non-local Fourier low \cite{ramu2015compact}. 
 
From the phenomenological point of view, we can directly apply the governing equation \eqref{HE} for the simulations of heat transfer in the powder bed fusion processes by using some effective properties $\kappa,\, c,\,\rho, \,g,\,\tau, \,l$ of the powder. Relation between these effective properties and the initial parameters of the  components in mixture theory (or in the two-temperature model) can be also derived \cite{forest2010some}.
In general, these material properties depend on the temperature, on the phase transition effects, etc. However, in the following analysis we will neglect such dependences like it is usually done in the simplified analytical approaches \cite{gladush2011physics}. Thus, we will use the averaged over the temperature range effective properties of the materials under consideration. 

Appropriate initial and boundary conditions of the considered model can be deduced based on the analysis of the balance equations or using the variational approach \cite{nguyen2005non,lurie2019variational}.

\subsection{Quasi-stationary problem}
\label{sec1b}
Considering problems of the moving heat sources we assume that the volumetric source $\hat q$ moves along $Ox$ axis with constant velocity $v$. It is convenient to define then a new coordinate system $O\xi yz$ that moves with the heat source such that:
\begin{equation}
\label{X}
\xi = x - vt
\end{equation}

Redefinitions of coordinates in the governing equation \eqref{HE} according to the standard rules ($\dot \theta \rightarrow - v \theta_{,\xi} + \dot\theta$,\, $\ddot\theta \rightarrow  v^2 \theta_{,\xi\xi} - 2v\dot\theta_{,\xi}+ \ddot\theta$,\, $\theta_{,x}\rightarrow\theta_{,\xi}$, etc.) reduce it to the following form:
\begin{equation}
\label{HExi}
(1 - g^2 \Delta)\dot \theta 
- v(1 - g^2 \Delta) \theta_{,\xi} 
- 2 \tau v \dot \theta_{,\xi} 
+ v^2 \tau \theta_{,\xi\xi}
+ \tau \ddot \theta
= \kappa (1 - l^2 \Delta) \Delta \theta + \frac{1}{c\rho}\hat q
\end{equation}

For the long enough bodies, the quasi-stationary condition can be achieved and the temperature distribution become independent on time. Corresponding steady-state form of equation \eqref{HExi} can be represented as follows: 
 \begin{equation}
\label{HEss} 
(1 - l^2 \Delta) \Delta \theta 
+ \frac{v}{\kappa}(1 - g^2 \Delta) \theta_{,\xi} 
- \frac{v^2 \tau}{\kappa}  \theta_{,\xi\xi}
+ \frac{1}{k}\hat q
= 0
\end{equation}
where $k$ is thermal conductivity coefficient ($\kappa= k /c\rho$) and comma denotes the differentiation with respect to the corresponding spatial coordinate. 

Note, that if we assume that the length scale parameters and the relaxation time equal to zero: $l=g=0$, $\tau=0$, then equation \eqref{HEss} reduces to classical equation that was initially used by Rosenthal and other researchers within the moving source problems \cite{rosenthal1946theory,panas2014moving}.

\subsection{Simplified gradient model and its general solution}
\label{sec1c}
Simulations with equation \eqref{HEss} or even with its initial transient variant \eqref{HE} can be performed by using numerical methods. Though, it is difficult to find a general solution and Green functions for these equations in a closed form. Using Hankel transform of Eq. \eqref{HEss}, it can be reduced to the fourth order ordinary differential equation, which can be easily solved, however the inverse transform cannot be performed analytically (such an approach have been used in Ref. \cite{ramu2015compact}). Direct application of standard Rosenthal's substitution $\theta(\xi,y,z) = \phi(\xi,y,z) e^{-\frac{u}{2\kappa}x}$ for the quasi-stationary problem \eqref{HEss} also does not make it easier to find the solution.

Thus, in the present work we propose to use some additional constitutive assumptions. This will allow us to obtain an approximate variant of the model that contains single additional length scale parameter and that can be resolved within the considered problems. Namely, we assume the following relations for the dissipation parameter and for the relaxation time:
 \begin{equation}
\label{as} 
g^2 =2l^2, \quad \tau = l^2/\kappa
\end{equation}

Substituting \eqref{as} into \eqref{HEss}, we obtain the following simplified variant of the model:
 \begin{equation}
\label{HEsss} 
(1 - l^2 \Delta) \Delta \theta 
+ \frac{v}{\kappa}(1 - 2l^2 \Delta) \theta_{,\xi} 
- l^2\frac{v^2}{\kappa^2}  \theta_{,\xi\xi}
+ \frac{1}{k}\hat q
= 0
\end{equation}

It is not obvious, why the solution of model \eqref{HEsss} can be more simple than those one of \eqref{HEss}. However, it can be checked by the direct substitution, that this equation \eqref{HEsss} can be reformulated in the following form with linear differential operators:
\begin{equation}
\begin{aligned}
\label{HEo} 
\mathcal H [\mathcal L [\theta]] + \frac{1}{k}\hat q = 0 \quad
\Longleftrightarrow \quad
(1 - l^2\Delta - l^2\frac{v}{\kappa}\partial_{\xi} )(\Delta\theta + \frac{v}{\kappa} \theta_{,\xi})
+ \frac{1}{k}\hat q = 0
\end{aligned}
\end{equation}
where 
\begin{equation}
\begin{aligned}
\label{HEod} 
\mathcal L [...] = \Delta + \frac{v}{\kappa} \partial_{\xi},\qquad
\mathcal H [...] = 1 - l^2 \mathcal L [...] = 1 - l^2\Delta - l^2\frac{v}{\kappa} \partial_{\xi} 
\end{aligned}
\end{equation}
where $\partial_{\xi}$ means the differentiation with respect to coordinate $\xi$.

Then, the solution of equation \eqref{HEsss} (or that is the same \eqref{HEo}) can be represented as the sum of classical part $\theta_0$ and additional gradient part $\theta_1$:
\begin{equation}
\label{GS} 
\theta = \theta_0 + \theta_1,\\
\end{equation}
wherein these parts of the solution obey the corresponding classical and modified gradient differential equations of the second order:
\begin{equation}
\label{GSo} 
\mathcal L[\theta_0] + \frac{1}{k}\hat q = 0\\ \qquad \text{and} \qquad
\frac{1}{l^2}\mathcal H[\theta_1] + \frac{1}{k}\hat q = 0
\end{equation}

Substituting solution \eqref{GS} into \eqref{HEo} and taking into account \eqref{GSo}, we can check that the governing equation of the model will be satisfied identically:
\begin{equation}
\begin{aligned}
\label{GSeq} 
\mathcal H [\mathcal L [\theta]] + \frac{1}{k}\hat q 
&= \mathcal H [\mathcal L [\theta_0 + \theta_1]] + \frac{1}{k}\hat q \\
&= \mathcal H [\mathcal L [\theta_0]] + \mathcal L [\mathcal H [\theta_1]]
+ \frac{1}{k}\hat q\\
&= -\mathcal H \left[\frac{1}{k}\hat q\right] 
- \mathcal L \left[l^2\frac{1}{k}\hat q\right] 
+ \frac{1}{k}\hat q\\
&= -\frac{1}{k}\hat q 
+ l^2 \mathcal L \left[\frac{1}{k}\hat q\right] 
- \mathcal L \left[l^2\frac{1}{k}\hat q\right] 
+ \frac{1}{k}\hat q \equiv 0
\end{aligned}
\end{equation}

Note, that solutions for equations \eqref{GSo} can be easily find based on the standard approaches used in the classical problems of moving heat sources. Thus, the representation \eqref{GS}, \eqref{GSo} give us a tool for construction of the general and particular solutions of the proposed simplified gradient model \eqref{HEsss}. Also we can use it to find the Green functions of the model.

Proposed approach for the development of the simplified models was initially used in gradient elasticity \cite{askes2011gradient}, in which the similar constitutive assumptions allow to obtain  the operator form of equilibrium equations with superposition of the classical elasticity operator and the Helmholtz or generalized Helmholtz operators \cite{askes2011gradient,lurie2006interphase}. Representation of the gradient elasticity solution through the gradient and classical parts (similar to \eqref{GS}) have been used, e.g. in micromechanics problems \cite{lurie2006interphase,solyaev2019three}.

\section{Gradient models of moving heat sources}
\label{sec3}
\subsection{Point and line heat sources models}
\label{intro}
\label{sec3a}
Point and line heat source models are usually used for the analysis of the so-called conductive and the key-hole modes of the melt pool formation, respectively \cite{patel2020melting}.
%Continuously acting moving point, line and plane heat 
Such type of the sources can be modeled by using the following spatial Dirac delta functions that are used as the right parts of the governing equations \cite{panas2014moving}:
\begin{equation}
\begin{aligned}
\label{RP} 
\hat q_{point} &= Q\, \delta(\xi)\delta(y)\delta(z) = Q\, \delta(\textbf r)\\
\hat q_{line} &= Q\, \delta(\xi)\delta(y)\\
%\hat q_{plane} &= Q\, \delta(\xi)
\end{aligned}
\end{equation}
where $\delta(...)$ is the Dirac delta function, $Q$ is the power magnitude and $\textbf r$ is the position vector in the moving coordinate system $O\xi yz$. 

Thus, we should find the solution of equation \eqref{HEo} for the volumetric power sources \eqref{RP} and we start with the case of the point heat source problem. Based on the proposed approach, we should solve the following equations to find classical $\theta_0$ and gradient $\theta_1$ parts of the solution:
\begin{equation}
\begin{aligned}
\label{AP1} 
\mathcal L[\theta_0] = - \frac{Q}{k} \delta(\textbf r) 
\quad\Longleftrightarrow\quad  
\Delta\theta_0 + \frac{v}{\kappa} (\theta_0)_{,\xi} = - \frac{Q}{k}\delta(\textbf r)
\end{aligned}
\end{equation}
\begin{equation}
\begin{aligned}
\label{AP2} 
\frac{1}{l^2}\mathcal H[\theta_1] = - \frac{Q}{k} \delta(\textbf r) 
\quad\Longleftrightarrow\quad  
\frac{1}{l^2}\theta_1 - \Delta\theta_1 - \frac{v}{\kappa} (\theta_1)_{,\xi} = 
- \frac{Q}{k} \delta(\textbf r)
\end{aligned}
\end{equation}

Equation \eqref{AP1} corresponds to the classical model of the moving point heat source and its solution is the widely known Rosental's formula \cite{rosenthal1946theory,panas2014moving}:
\begin{equation}
\begin{aligned}
\label{AP1sol} 
\theta_0 = \frac{Q}{4\pi k R} \text{e}^{-\frac{v}{2\kappa}(\xi + R)}
\end{aligned}
\end{equation}
where $R=|\textbf r| = \sqrt{\xi^2+y^2+z^2}$.

Modified gradient equation \eqref{AP2} is not more complex than the classical one and it can be also solved following the approach proposed by Rosenthal \cite{rosenthal1946theory,panas2014moving} for the derivation of classical solution \eqref{AP1sol}. Therefore, we assume that the gradient part of the solution can be represented as follows:
\begin{equation}
\begin{aligned}
\label{AP2as} 
\theta_1(\textbf r) = \phi(\textbf r) \text e^{-\frac{v}{2\kappa} \xi}
\end{aligned}
\end{equation}

Substituting \eqref{AP2as} into \eqref{AP2} we can obtain the following Helmholtz-type equation with respect to function $\phi(\textbf r)$:
\begin{equation*}
\frac{1}{l^2}\phi 
- \Delta\phi + \underline{\frac{v}{\kappa} \phi_{,\xi}} - \frac{v^2}{4\kappa^2}\phi
- \underline{\frac{v}{\kappa} \phi_{,\xi}} + \frac{v^2}{2\kappa^2} \phi = 
- \frac{Q}{k} \delta(\textbf r) \text e^{\frac{v}{2\kappa} \xi}
\end{equation*}
\begin{equation}
\label{AP2f} 
\Longrightarrow\quad \Delta\phi - \left(\frac{1}{l^2} + \frac{v^2}{4\kappa^2}\right)\phi
= \frac{Q}{k} \delta(\textbf r)
\end{equation}
where it is seen that the underlined terms are canceled and we take into account that $\delta(\textbf r) \text e^{\frac{v}{2\kappa} \xi} = \delta(\textbf r)$ (see Ref.\cite{eagar1983temperature}).

The main advantage of substitution \eqref{AP2as} is that it allows to avoid the first order derivative of the function (corresponding terms are cancelled) and to reduce the initial differential equation to the standard radially-symmetric Helmholtz equation. Note, that this assumption \eqref{AP2as} can be effectively used only in the framework of presented simplified variant of gradient theory with relations between additional gradient parameters given by \eqref{as}. In the case of general model \eqref{HEss}, or in the case of the models with some other type of constitutive assumptions instead of \eqref{as}, the substitution \eqref{AP2as} will not work, i.e. the first order and moreover, the second order derivatives  of $\phi(\textbf r)$ will remain in the final equation. This will make further difficulties for the analytical solution. Thus, the physical meaning of the constitutive assumptions \eqref{as} is that there exists some similarity between the gradient and the classical parts of the solution, such that the effects of the heat source movement can be represented by the same decaying function $\text e^{-\frac{v}{2\kappa} \xi}$ and the rest part of the problem can be reduced to the solution of the Helmholtz equations. 

Based on \eqref{AP2f} and \eqref{AP2as} we find then the gradient part of the solution as follows:
\begin{equation}
\label{AP2sol} 
\begin{aligned}
\phi(\textbf r) &= - \frac{Q}{4\pi k R} \text{e}^{-\sqrt{\frac{1}{l^2}+\frac{v^2}{4\kappa^2}}R}\\[5pt]
\theta_1(\textbf r) &= - \frac{Q}{4\pi k R} \text{e}^{-\frac{v}{2\kappa} \xi-\sqrt{\frac{1}{l^2}+\frac{v^2}{4\kappa^2}}R}
\end{aligned}
\end{equation}

and the total solution for the temperature distribution $\theta=\theta_0+\theta_1$ becomes to:

\begin{equation}
\begin{aligned}
\label{GFp} 
\theta(\textbf r) &= \frac{Q}{4\pi k} \frac{\text{e}^{-\frac{v}{2\kappa}\xi}}{R}
\left(
\text{e}^{-\frac{v}{2\kappa}R} - 
\text{e}^{-\sqrt{\frac{1}{l^2}+\frac{v^2}{4\kappa^2}}R}
\right)
\end{aligned}
\end{equation}

For the powder bed fusion processes, the problem of the heat source that moves over the half-space is of interest. Solution of this problem can be approximated based on the doubled solution for the infinite space assuming the thermal insulation condition at the free surface (i.e. neglecting the convective and radiative heat transfer like it is usually done in similar classical models \cite{panas2014moving}). Then, the dimensionless form of the solution for the point source that moves over the half-space can be defined based on \eqref{GFp} as follows:
\begin{equation}
\begin{aligned}
\label{GFpn} 
\bar \theta_p(\bar{\textbf r}) &= n \frac{\text{e}^{-\bar \xi}}{\bar R}
\left(
\text{e}^{-\bar R} - 
\text{e}^{-\sqrt{1+\, \text{Pe}_m^{-2}}\,\bar R}
\right)
\end{aligned}
\end{equation}
where $\bar \theta_p(\bar{\textbf r}) = \theta(\bar{\textbf r})/(T_m-T_i)$ is the rise of temperature normalized with respect to some critical value that can be related, e.g. with the material melting point $T_m$; 
$n = \frac{Q v }{4\pi \kappa^2 \rho c(T_m-T_i)}$ is the so-called operating parameter \cite{eagar1983temperature}; $\bar \xi = \frac{v \xi}{2\kappa}$, $\bar R = \frac{v R}{2\kappa}$,  $\bar{\textbf r} = \frac{v}{2\kappa}\textbf r$ are the non-dimensional coordinates.

The main feature of solution \eqref{GFpn} is the presence of additional non-dimensional group of parameters that we defined here as the micro-scale Peclet number:
\begin{equation}
\label{pem} 
\text{Pe}_m = \frac{v l}{2\kappa}
\end{equation}

In classical models of heat and mass transfer, the value of Peclet number (Pe) defines the ratio between the rate of convection and the rate of diffusion processes. In application to the moving heat source problems, Peclet number is introduced usually to define the ratio of the heat diffusion characteristic time $L^2/\kappa$ to the transit time of the heat source $2L/v$, i.e. Pe $=vL/(2\kappa)$, where $L$ is some macroscopic characteristic length scale of the problem \cite{hou2000general}. 

In the present case in \eqref{GFpn},\eqref{pem}, we have the definition of the micro-scale Peclet number through the internal length scale parameter of material $l$. This additional parameter of gradient theory defines the intensity of non-classical effects in the gradient solutions and amount of corrections that can be obtained for the classical predictions for the temperature distribution and, e.g. for the melt pool shape and size. 
In the case of its small value ($l \rightarrow 0$ and $\text{Pe}_m \rightarrow 0$) we obtained the classical theory and classical Rosenthal solution for the moving heat source \cite{rosenthal1946theory,panas2014moving}. In the case of large values of $l$ and $\text{Pe}_m$ the gradient effects become pronounced. 

Note, that such non-dimensional parameter $\text{Pe}_m$ is specific for the heat transfer problems. For example, in the inclusion problems of gradient elasticity and gradient hydrodynamics corresponding non-dimensional group is defined as the ratio of inclusion size and the length scale parameter \cite{solyaev2019three}. In wave propagation problems, the length scale parameter arises in gradient solutions as the ratio with the wavelength \cite{solyaev2021electric}.

In opposite to classical Rosenthal's solution, gradient solution \eqref{GFpn} does not contain singularity and the maximum value of the temperature rise is given by:
\begin{equation}
\begin{aligned}
\label{GFpmax} 
\bar \theta_{p,max} = 
\lim\limits_{\textbf r \rightarrow 0}\bar \theta_p = n \left(  
\sqrt{1+\, \text{Pe}_m^{-2}} - 1\right)
\end{aligned}
\end{equation}

Regularization of classical singular solutions is typical for gradient theories \cite{lazar2020second,lazar2013fundamentals}. In the present case this formula \eqref{GFpmax} may not be such an effective for the description of the real fusion processes, since in the area of maximum heating there arise additional effects related to melting, partial evaporation, hydrodynamics of melt pool, etc. However, the assessment for the maximum temperature \eqref{GFpmax} can be useful when the heat input power is not very high, e.g. in some laser treatment applications. Also we can use Eq. \eqref{GFpmax} to identify  the value of the material's length scale parameter. This can be done in the following way. From the experiments one can find the minimum value of the operating parameter $n$ that is needed for the occurrence of melting (see, e.g. Ref.\cite{li2004comparison}).  This situation corresponds to the case, when the normalized temperature rise equals to one, i.e. $n_{min} = n: \,\, \bar \theta_{p,max}=1$.
Then, based on Eq. \eqref{GFpmax} we can identify the value of the micro-scale Peclet number as follows:
\begin{equation}
\begin{aligned}
\label{Pei} 
\text{Pe}_m = \frac{n_{min}}{\sqrt{1+2n_{min}}}
\end{aligned}
\end{equation}

Therefore, the classical case with Pe$_m=0$ corresponds to zero value of minimum operating parameter $n_{min}=0$, i.e. for any infinitesimal external heat input applied in a single (moving) point there will arise some melted area. In opposite, the gradient model assumes that material can sustain some amount of concentrated heat input without melting.  

Returning to the dimensional definitions in \eqref{Pei} after some algebraic simplifications we can obtain a closed form assessment for the experimental identification of the material's length scale parameter:

\begin{equation}
\begin{aligned}
\label{li} 
l = \frac{Q_{min}}{2\pi k(T_m-T_0) \sqrt{1+\frac{Q_{min}v}{2\pi k (T_m-T_0) \kappa}}}
\end{aligned}
\end{equation}
where $Q_{min} = \lambda P_{min}$ is the power magnitude; $\lambda$ is the absorptivity of the material; $P_{min}$ is the minimum laser power needed for the occurrence of melting in a given material with melting point $T_m$ and with properties $k$ and $\kappa$ (averaged over the  temperature range $[T_i,T_m]$) in the case of the laser scanning speed $v$.

Stability of the identified values of parameter $l$ from the experiments with different scanning speed $v$ and corresponding values of $P_{min}$ can be used for validation of the presented gradient model. Solution for the stationary sources can be also easily obtained from the developed solution assuming that $v=0$.

Based on the solution \eqref{GFpn} we can also evaluate the cooling rate that is realized around the moving heat source. This quantity can be related to the thermal gradient in the motion direction as follows \cite{cline1977heat}:
\begin{equation}
\begin{aligned}
\label{GFpmaxcr} 
\frac{\partial \bar \theta_p}{\partial t} = - v \frac{\partial \bar \theta_p}{\partial \bar\xi} 
= v \left( \left(1+\frac{\bar\xi}{\bar R^2}+\frac{\bar\xi}{\bar R}\right) \bar \theta_p(\bar{\textbf r}) 
+  \frac{\bar\xi}{\bar R}
\left(\sqrt{1+\text{Pe}_m^{-2}} - 1\right) \bar \theta_1(\bar{\textbf r})
\right)
\end{aligned}
\end{equation}
where $\bar \theta_1$ is the non-dimensional gradient part of the solution \eqref{AP2sol}.\\

Considering problem with the line source $\hat q_{line}$ (see \eqref{RP}) we can reduce it to the two-dimensional statement and solve in a polar coordinates. Solution of this problem is very similar to the previous one and we give it without derivations:
\begin{equation}
\begin{aligned}
\label{GFl} 
\theta_l(\textbf r) &= \frac{Q}{2\pi k} \text{e}^{-\frac{v}{2\kappa}\xi}
\left(
K_0\left(\frac{v r}{2\kappa}\right) - 
K_0\left(r\,\sqrt{\frac{1}{l^2}+\frac{v^2}{4\kappa^2}}\right)
\right)
\end{aligned}
\end{equation}
where $K_0$ is zero order modified Bessel function of the second kind and $r=\sqrt{\xi^2+y^2}$ is the radial distance from the line source to the given point in polar coordinates.

Dimensionless form of solution \eqref{GFl} is given by:
\begin{equation}
\begin{aligned}
\label{GFln} 
\bar \theta_l(\bar{\textbf r}) &= n \, \text{e}^{-\bar \xi}
\left(
K_0\left(\bar r\right) - 
K_0\left(\bar r\,\sqrt{1+\text{Pe}_m^{-2}}\right)
\right)
\end{aligned}
\end{equation}

As previously, reduction to classical solution $\bar \theta_{l,class}(\bar{\textbf r}) = n\, \text{e}^{-\bar \xi}K_0\left(\bar r\right)$ \cite{panas2014moving} is realized for the case of zero value of Pe$_m$ number. Maximum temperature rise is also finite in this gradient solution \eqref{GFln}, however the location of this maximum shifts out from the origin of coordinate system (this will be illustrated in the Results section). Approximate value of the maximum temperature rise for a not very high values of Pe$_m$ can be obtained based on the series expansion of \eqref{GFln} at the origin of coordinate system and it is given by:
\begin{equation}
\begin{aligned}
\label{GFlns} 
\bar \theta_{l,max}&= \frac{n}{2} \log{\left(1+\text{Pe}_m^{-2}\right)}
\end{aligned}
\end{equation}

\subsection{Gaussian heat source}
Point source model is a rough approximation for the real distribution of a heat flux around the laser spot. More realistic will be the models of circular heat sources with Gaussian distribution of a heat flux \cite{eagar1983temperature, yan2018review}. Solutions for temperature rise in such models can be found based on the Green function method. Appropriate Green function for the considered model can be obtained based on the point source solution \eqref{GFp}  using similar to classical approach that was described, e.g. in Refs. \cite{eagar1983temperature, panas2014moving}. One should consider the problem with two point sources placed symmetrically with respect to the plane $z=0$. Assuming that distance between these sources tends to zero, one can obtain the following expression for the Green function based on \eqref{GFp}:
\begin{equation}
\begin{aligned}
\label{GFpc} 
G(\textbf r,\textbf r_0) &= \frac{1}{2\pi k } 
\frac{
\text{e}^{-\frac{v}{2\kappa}(\xi-\xi_0)}}{R_0}
\left(
\text{e}^{-\frac{v}{2\kappa}R_0} - 
\text{e}^{-\sqrt{\frac{1}{l^2}+\frac{v^2}{4\kappa^2}}\,R_0}
\right)
\end{aligned}
\end{equation}
where $R_0=|\textbf r-\textbf r_0| = \sqrt{(\xi-\xi_0)^2+(y-y_0)^2+z^2}$ 
is the distance between actual point under consideration $\textbf r=\{\xi,y,z\}$ and the location of the point source on the surface of the half-space $\textbf r_0=\{\xi_0,y_0,0\}$; and the magnitude of the power input is $Q=1$.
Classical Green function follows from Eq. \eqref{GFpc} in the case $l\rightarrow0$.

The energy distribution of the Gaussian laser beam is defined by \cite{eagar1983temperature}:
\begin{equation}
\begin{aligned}
\label{gaus} 
\hat q_g(\xi,y) = \frac{Q}{2\pi a^2 }\, 
\text e^{-\frac{\xi^2+y^2}{2a^2}}
\end{aligned}
\end{equation}
where $\lambda$ is the absorptivity of the material and $a$ is the characteristic laser beam radius.

Then, the temperature rise can be evaluated through the convolution of the Green function \eqref{GFpc} and the distribution \eqref{gaus}. Final solution can be presented in the following dimensionless form in cylindrical coordinate system $\textbf r = \{r, \phi, z\}$:
\begin{equation}
\begin{aligned}
\label{conv} 
&\bar\theta_g(\bar{\textbf r})= 
\int\limits_{0}^{\bar r_{max}} 
\int\limits_{0}^{2\pi}
\hat q_g(\bar\xi_0, \bar y_0) \,G(\bar{\textbf r},\bar{\textbf r}_0) \,\bar r_0\,
d\phi_0d\bar r_0\\
& =
\frac{n}{\pi \text{Pe}^{2}}
\int\limits_{0}^{\bar r_{max}} 
\int\limits_{0}^{2\pi}
\frac{
\text{e}^{-(\bar r\cos\phi-\bar r_0\cos\phi_0)
- \bar r_0^2/(2\text{Pe}^{2})}}{\bar R_0}
\left(
\text{e}^{-\bar R_0} - 
\text{e}^{-\sqrt{1+\text{Pe}_m^{-2}}\,\bar R_0}
\right)
\bar r_0
d\phi_0d\bar r_0
\end{aligned}
\end{equation}
where 
$\bar r_0 = \frac{v}{2\kappa} |\textbf r_0| =\frac{v}{2\kappa} \sqrt{\xi_0^2+y_0^2}$ is the non-dimensional radial coordinate of the point heat source; and Pe $=\frac{va}{2\kappa}$ is the standard definition for the Peclet number of the problem.

In general case, upper limit of integration along radial coordinate in \eqref{conv} should be $\bar r_{max}=\frac{v}{2\kappa}r_{max}=\infty$, however, we can take into account that the energy of Gaussian laser beam is located mostly inside the circular area which radius equals to $5a$ \cite{li2004comparison}. Then, in numerical calculations we can use the non-dimensional upper limit defined by $\bar r_{max}= 5$Pe.

Peculiarity of this integral is that it does not contain singularities and it can be evaluated numerically even more easily than the similar one in classical models (see, e.g. Ref.\cite{li2004comparison}). This integral contains the macro-scale and the micro-scale dimensionless parameters Pe and Pe$_m$. The relation between these parameters defines the ratio between the beam radius and the length scale parameter: Pe/Pe$_m = a/l$. And it is notable, that we cannot simplify the dimensionless form of the solution such that it contains the single non-dimensional group of parameters.

\section{Results and discussion}
\label{res}
In this section we give the illustrations for the derived solutions. Also we give an example of identification of the length scale parameter of the model. All calculations are performed with the dimensionless forms of the solutions. Unless otherwise stated, we use the unit value of the operating parameter $n=1$. 

\begin{figure}[!t]
\textbf a
\includegraphics[width=0.45\linewidth]{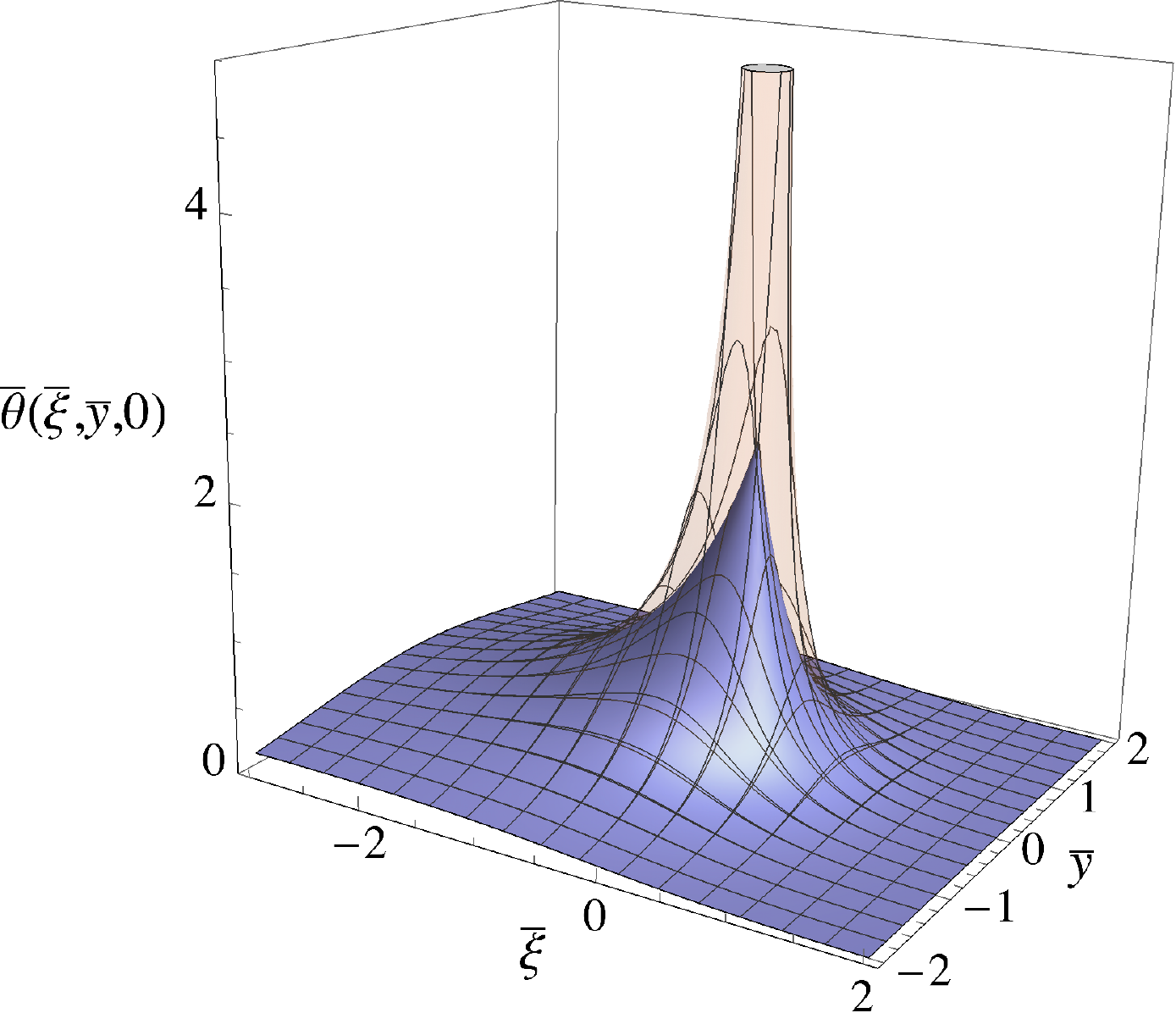}\quad
\textbf b
\includegraphics[width=0.45\linewidth]{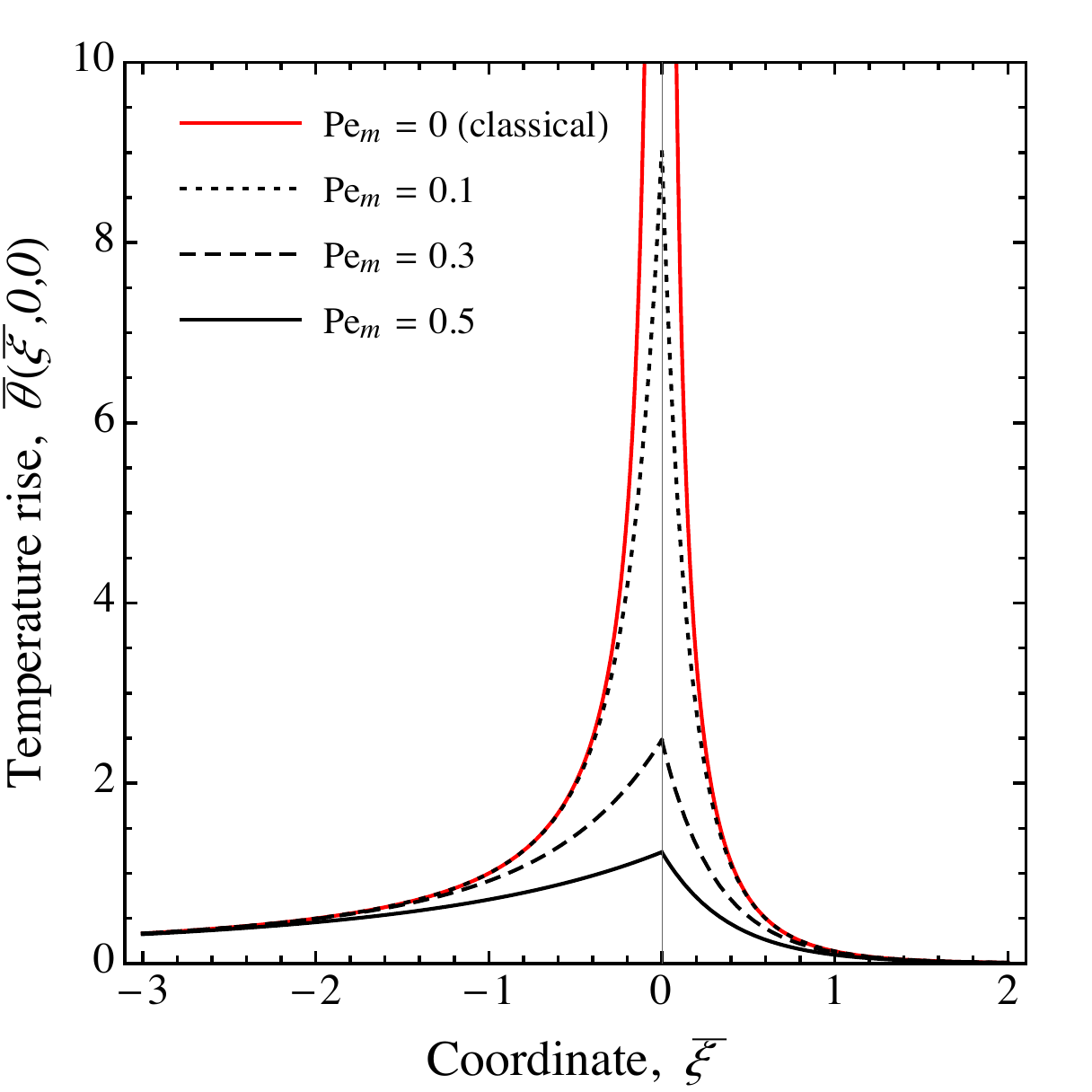}\\[10pt]
\textbf c
\includegraphics[width=0.45\linewidth]{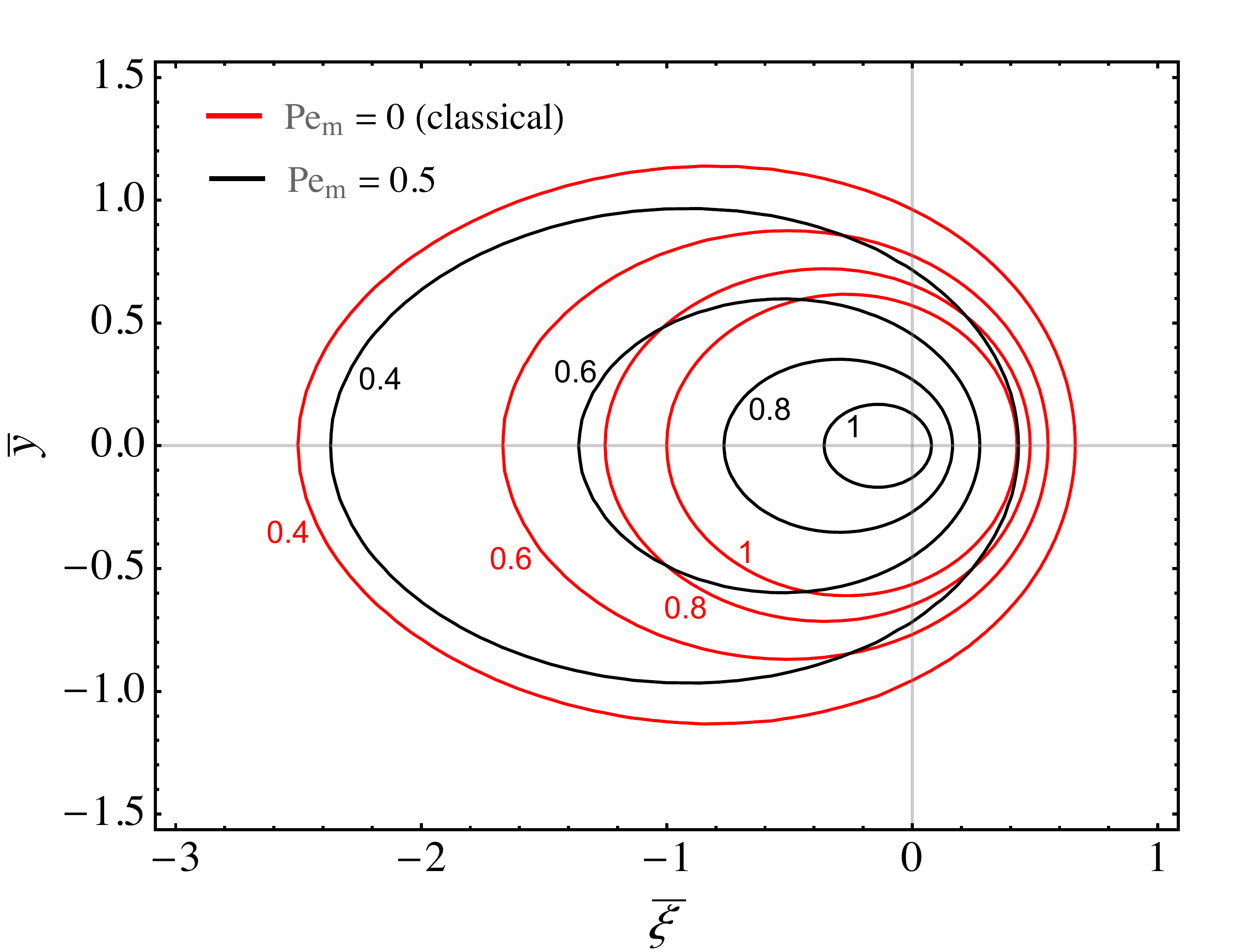}\quad
\textbf d
\includegraphics[width=0.45\linewidth]{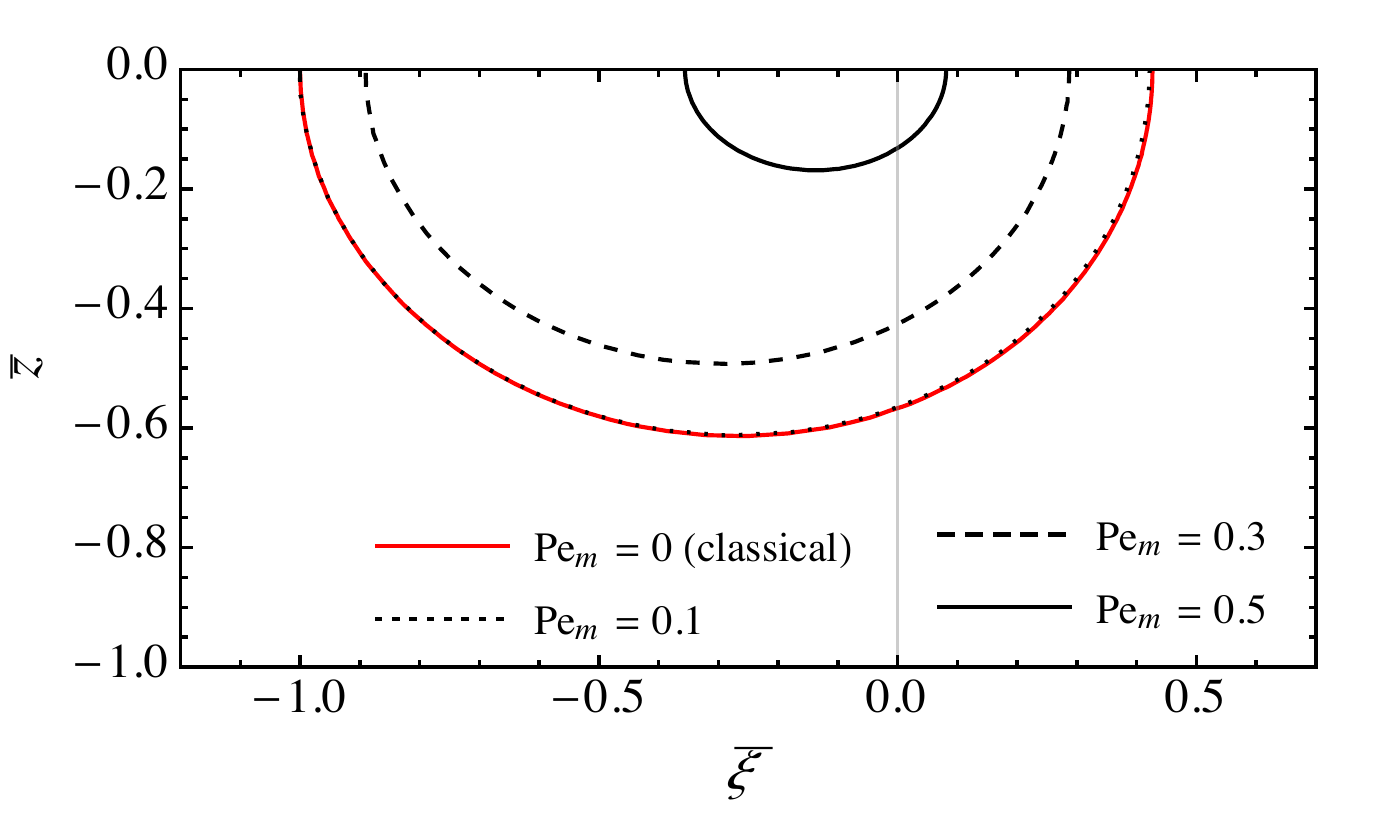}
\caption{ \textbf a: Distribution of dimensionless temperature rise in classical (transparent color) and gradient (blue color) solutions due to action of moving point heat source at the half-space surface $\bar z=0$, \textbf b: Influence of the micro-sale Peclet number on the temperature profile along the direction of movement, \textbf b: Isotherms in the heat affected zone at the surface $\bar z=0$, \textbf d: change of the melt pool shape with increase of the micro-scale Peclet number.}
\label{fig1}
\end{figure}

Comparison between classical and gradient \eqref{GFpn} solutions for the moving point heat source is presented in Fig. \ref{fig1}. Here we present the distribution for the temperature rise at the surface $z=0$ (Fig. \ref{fig1}a, b). It is seen, that in comparison with standard Rosenthal's solution, the gradient solution predicts lower temperature rise around the action of point heat source. Far from this point, both solutions have similar asymptotic behavior and the temperature profiles almost coincide at the distance $\bar\xi<-2$ behind the point source (Fig. \ref{fig1}b). In gradient solution, there arise a fast decrease of the maximum temperature rise $\bar\theta_{p,max}$ with increase of the micro-scale Peclet number Pe$_m$ (Fig. \ref{fig1}b, black lines). In accordance to Eqv. \eqref{GFpmax}, the value of $\bar\theta_{p,max}$ reduces from the infinity (classical solution, red line in Fig. \ref{fig1}b) to the unit value when the non-dimensional number Pe$_m$ changes from 0 to $\sim$0.5.

The shape of the heat affected zone and of the melt pool are also significantly affected by the gradient effects. In Figs. \ref{fig1}c, d we present the comparison between the  isotherms in classical and gradient solutions at the surface of the half-space $\bar z =0$ and at the sagittal plane $\bar y =0$, respectively. It is seen, that the width and the length of the heat affected zone become smaller in gradient solution (Fig. \ref{fig1}c). The melt pool shape corresponds to the isotherm $\bar\theta=1$ in Figs. \ref{fig1}c, d. It is seen, that the melt pool size reduces with the increase of the micro-scale number (Fig. \ref{fig1}b). Thus, when the material's internal characteristic length scale $l$ becomes large enough, the solution predicts the considerable decrease of the melt pool dimensions. The value of the melt pool depth $\bar z_{max}$ can be evaluated based on the derived solutions as the maximum value of coordinate $\bar z$  that corresponds to the isotherm $\bar\theta=1$ in the sagittal plane $\bar y =0$, i.e.:
\begin{equation}
\begin{aligned}
\label{zmax} 
\bar z_{max} = \{\max |\bar z|: \bar\theta(\bar \xi,0,\bar z) = 1\}
\end{aligned}
\end{equation}

\begin{figure}
\textbf a
\includegraphics[width=0.3\linewidth]{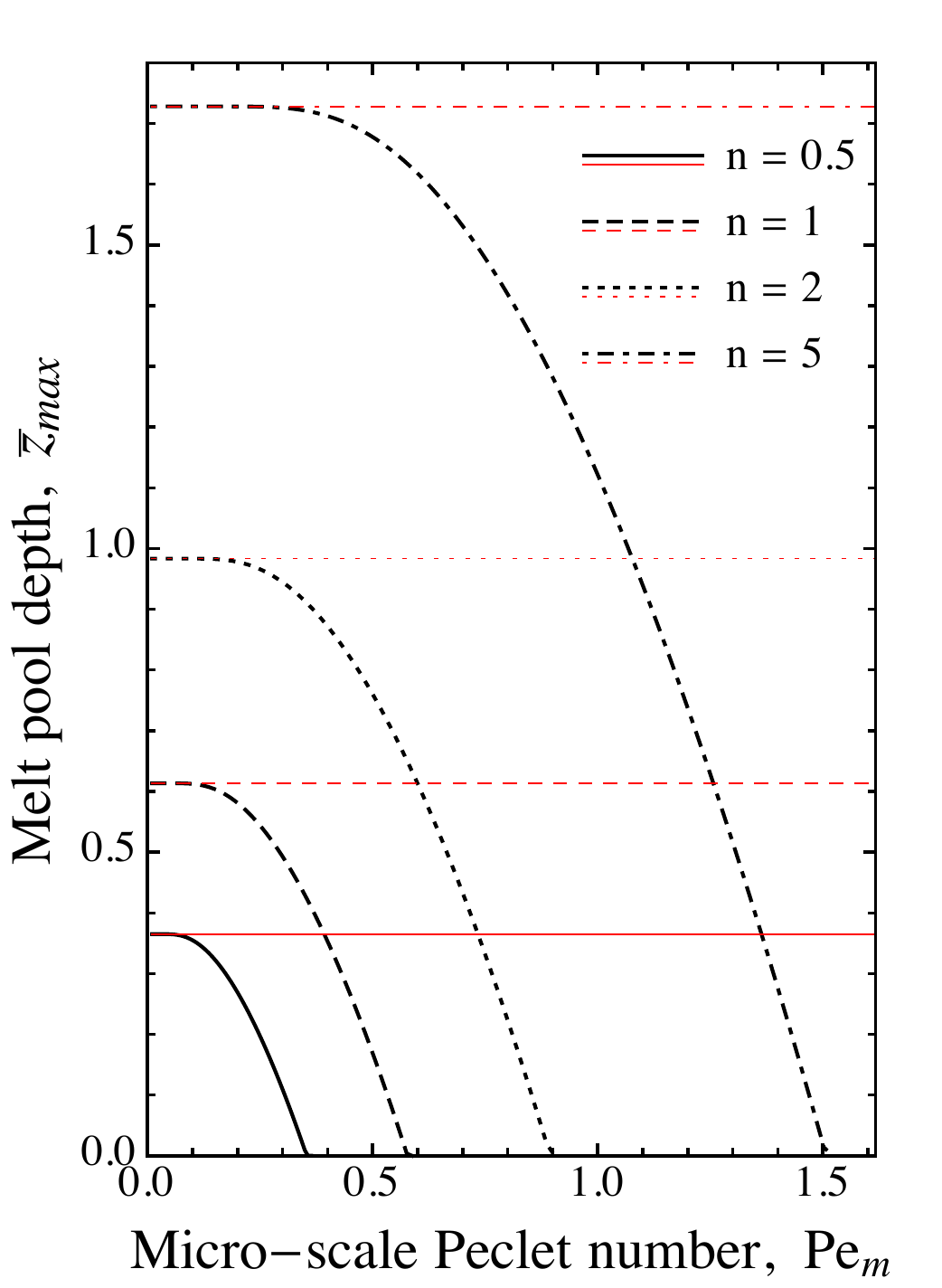}
\textbf b
\includegraphics[width=0.3\linewidth]{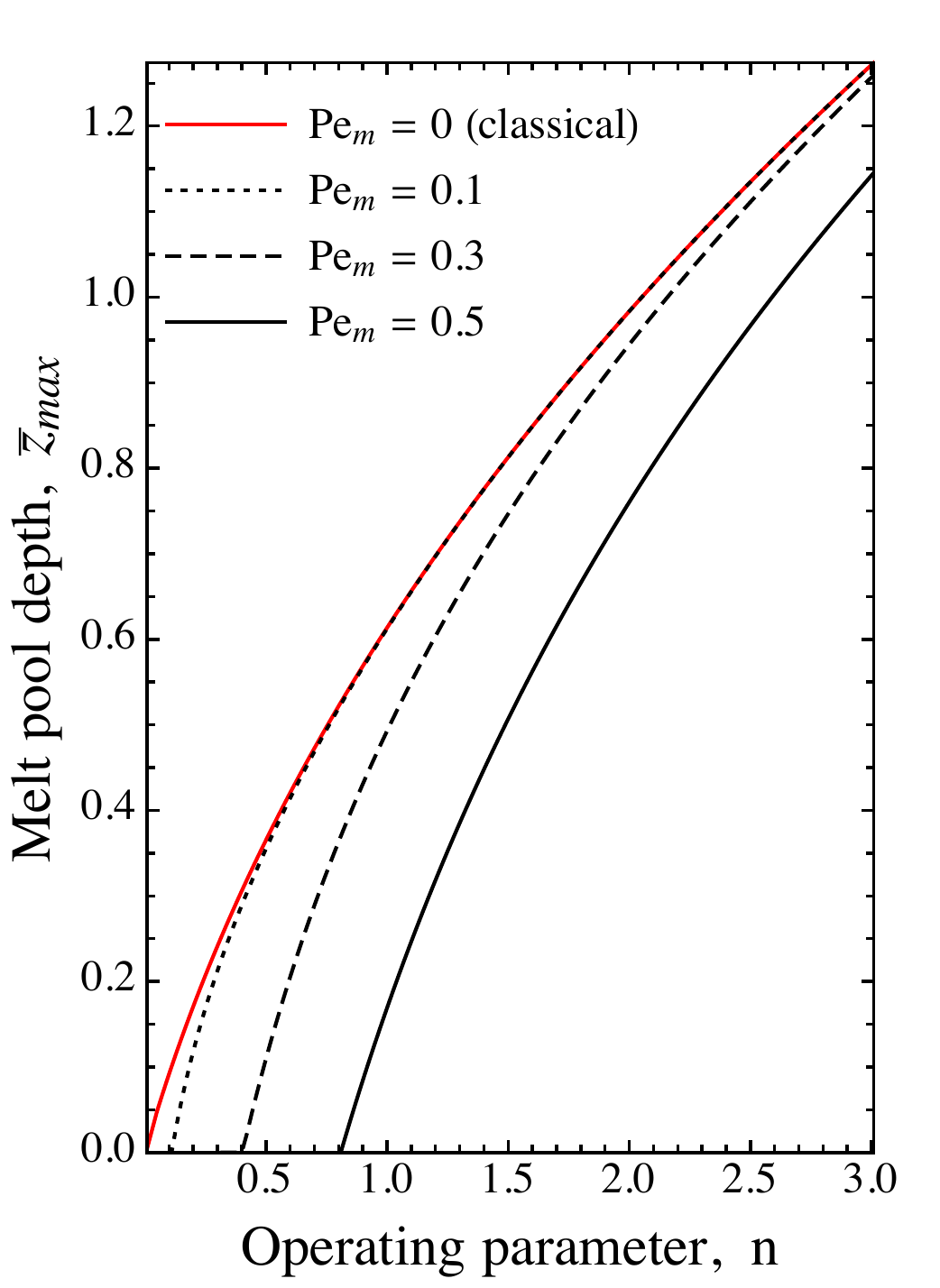}
\textbf c
\includegraphics[width=0.31\linewidth]{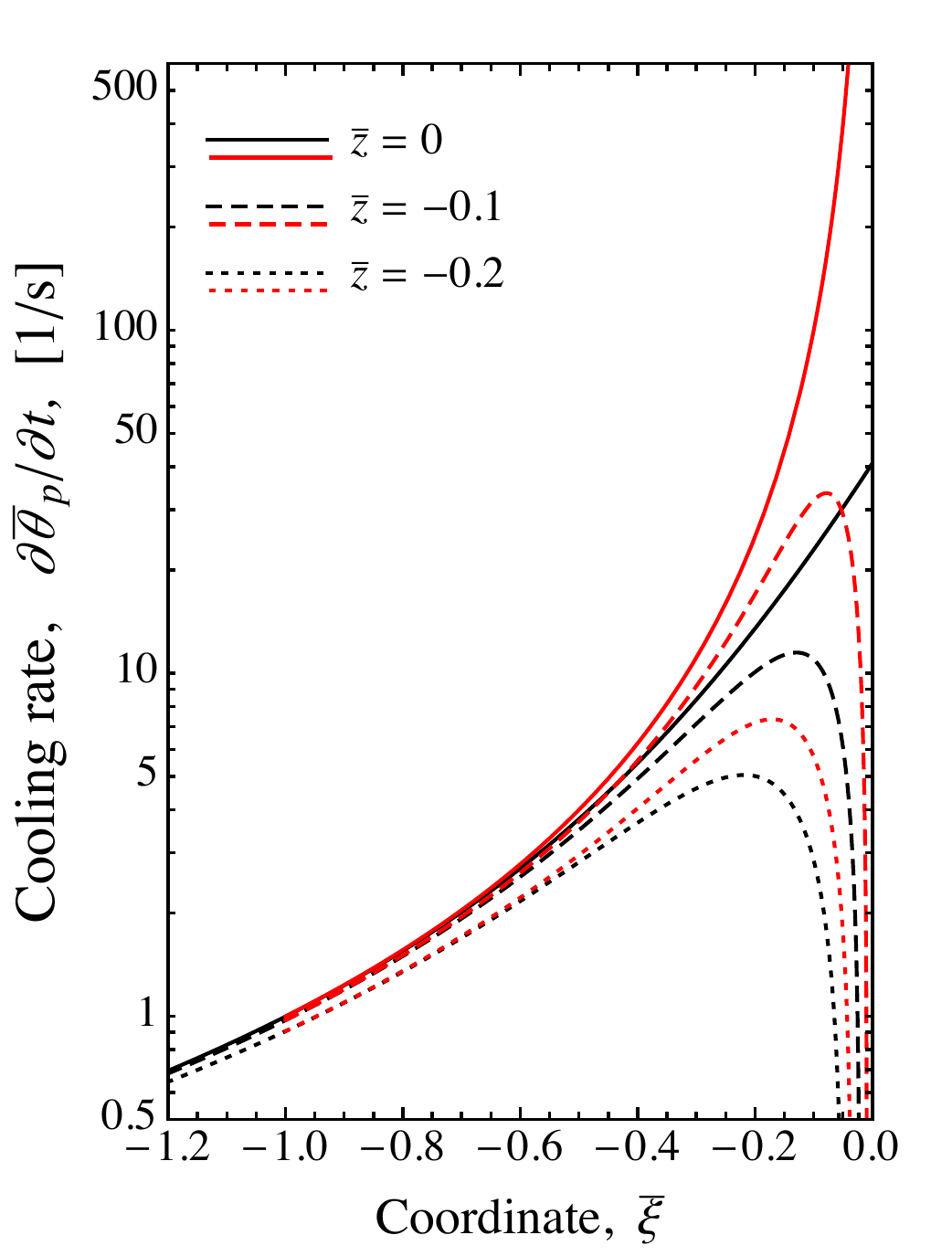}
\caption{Dependence of the melt pool depth on the micro-scale Peclet number (\textbf a) and on the operating parameter (\textbf b); and the cooling rate distribution behind the heat source at different depths $\bar z$ (\textbf c). Red and black colors correspond to classical and gradient solutions, respectively, in these plots.}
\label{fig2}
\end{figure}

This problem \eqref{zmax} can be solved numerically. For the classical solution corresponding analytical approximations have been also presented \cite{ramos2019analytical}. In Fig. \ref{fig2} we show the evaluated dependences of the melt pool depth on the micro-scale Peclet number and on the operating parameter $n$. It is seen, that with increase of Pe$_m$  the melt pool depth reduces down to zero, while for small values of Pe$_m$ gradient solution asymptotically approaches some classical values (Fig. \ref{fig2}a). With increase of the operating parameter the depth of melt pool become larger. However, there exist some ranges of the heat input power (operating parameter defines this power in the dimensionless solution) when the melt pool does not arise (Fig. \ref{fig2}b). And in opposite to classical solution for the point heat source, in the gradient solution the dependence $\bar z (n)$ starts from some non-zero value and the material main sustain some amount of the concentrated heat input without melting (Fig. \ref{fig2}b, black curves). Corresponding minimum value of the heat source power for the given value of the micro-scale Peclet number is defined by the equation \eqref{Pei}. 

Thus, from the experiments with laser melting of the powder materials one may measured the dependence of the melt poole depth on the operating parameter and overlay the obtained experimental data with the plots similar to those one presented in Fig. \ref{fig2}b. In such a way the presented model can be validated and the length scale parameter of the model can be identified. For the welding of solid materials such experiments have been widely performed and the models of moving heat sources with different distributions of heat flux over some finite size area (as the generalization of the point source model) were validated \cite{eagar1983temperature}. However, for the solid materials the influence of the gradient effect may not be such pronounced. We suppose that the most significant clarifications within the gradient theories can be obtained for the fusion processes of the powder materials, in which the microstructural effects play a significant role \cite{barchiesi2021granular}. 

Another approach for validation of the presented gradient models can be related to the measuring of the cooling rate during the laser melting of the powders. In Fig. \ref{fig2}c it is shown that the gradient solution \eqref{GFpmaxcr} predicts the finite level of the cooling rate at the point of action of the heat source. Note, that the surface and subsurface cooling rates can measure experimentally 
\cite{pauly2018experimental}. Thus, the simplified gradient model even for the point source can be used for the processing of the corresponding experimental data.
%on the cooling rates. The maximum value of the cooling rate can be also related to the length scale parameter of the model.
%in the laser powder bed fusion.

%\cite{pauly2018experimental} or the subsurface \cite{thampy2020subsurface} cooling rates in laser powder bed fusion. 
%Similar classical solution cannot be applied in such case, since it predicts the infinite level of the cooling rate at the surface (Fig. \ref{fig2}c, red color). More complex numerical or semi-analytical models with different distribution of heat flux should be involved within the classical heat transfer theory for such processing. Thus, the closed form solutions provided by gradient theories could be more useful in such applications.

\begin{figure}
\textbf a
\includegraphics[width=0.4\linewidth]{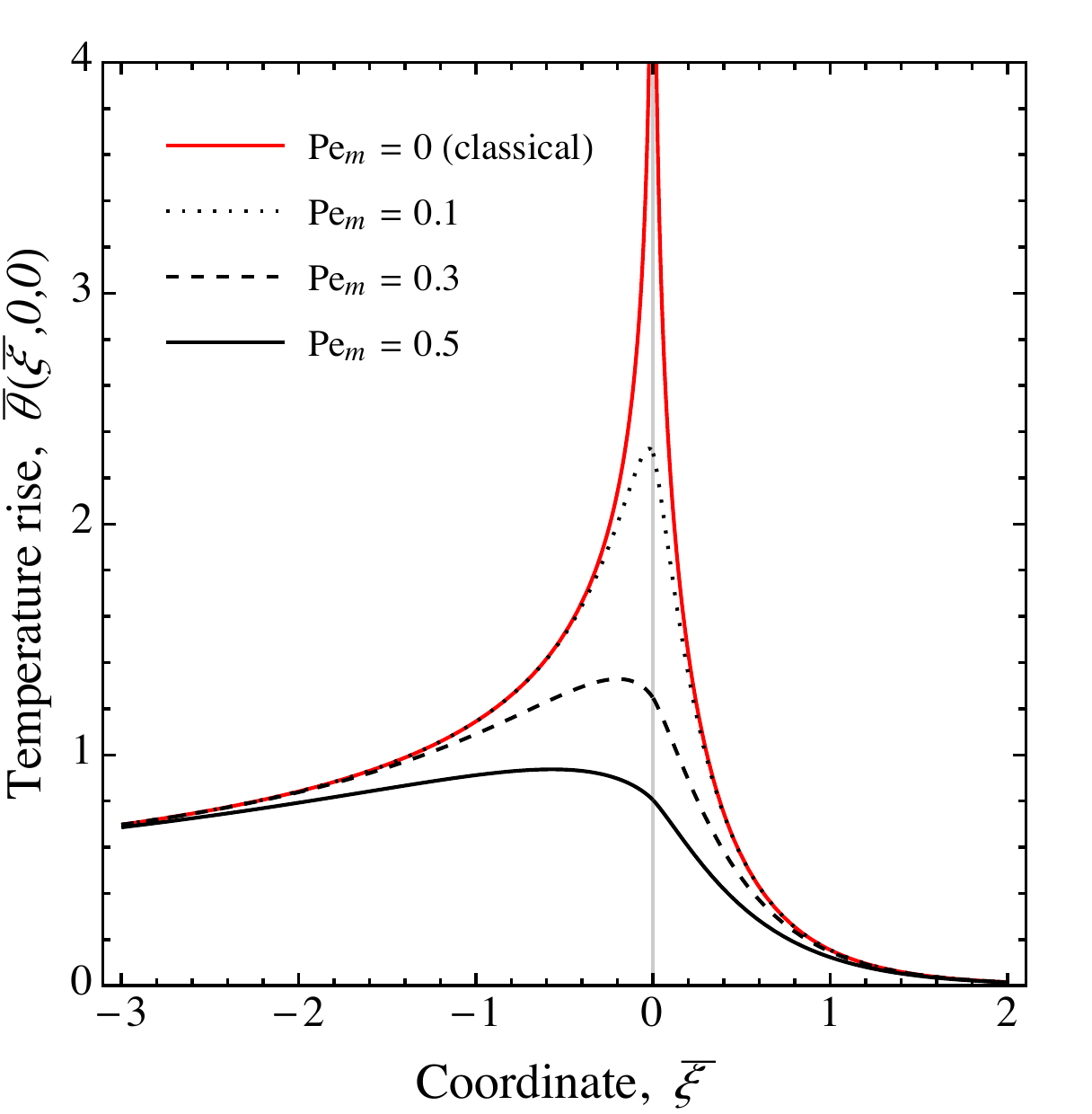}\,\,
\textbf b
\includegraphics[width=0.5\linewidth]{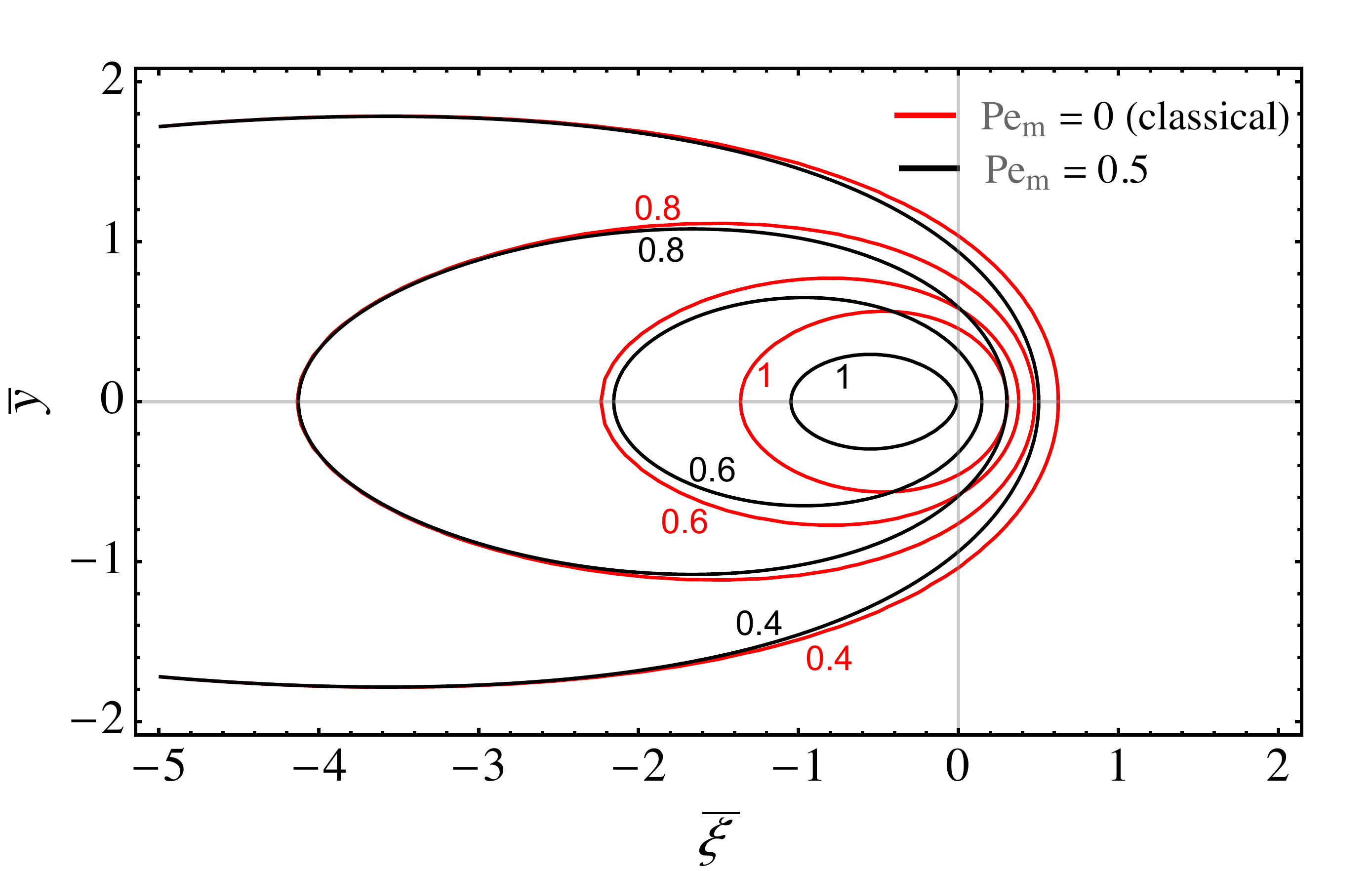}
\caption{Illustrations for the two-dimensional solution of the moving line heat source problem, \textbf a: temperature rise along the movement direction, \textbf b: isotherms of the solution}
\label{fig3}
\end{figure}

Gradient solution for the moving line  source is illustrated in Fig. \ref{fig3}. Here it is seen, that in opposite to classical solution, the maximum temperature rise arise not at the centre of the coordinate system, but with some shift in opposite to the direction of movement (Fig. \ref{fig3}, black lines). This is a typical situation, that arises even in classical solutions for the problems with Gaussian heat flux distribution (see below) and other type of the more general heat source models\cite{hou2000general}. Assessment for the maximum temperature rise in gradient solution was given for the line heat source by equation \eqref{GFpmax} that was evaluated at the origin of the coordinate system. From Fig. \ref{fig3}a it is seen, that this assessment is rather accurate and even in the case of the high gradient effects (large Pe$_m$) the maximum temperature rise is closed to those one at the origin of coordinates.  

Another peculiarity of the line heat source solution can be seen in the Fig. \ref{fig3}b where we plot the isotherms at the plane $\bar\xi$-$\bar y$. It is shown, that in opposite to the point source solution (Fig. \ref{fig1}d), the gradient effects do not make such a strong influence on the line source isotherms. All effects are concentrated in the melt pool zone (contour line $\bar\theta=1$) and for the smaller levels of heating ($\bar\theta\leq0.8$) the isotherms become close to classical. The line source model is used usually for the simulation of the high-power processes with the key-hole mode formation of the melt pool. For such processes the gradient clarifications become less important in the area outside from the melt pool. Nevertheless, inside the melt pool region it can be used for the refined assessments for its shape, cooling rates, etc.

\begin{figure}[b!]
\centering
\includegraphics[width=0.45\linewidth]{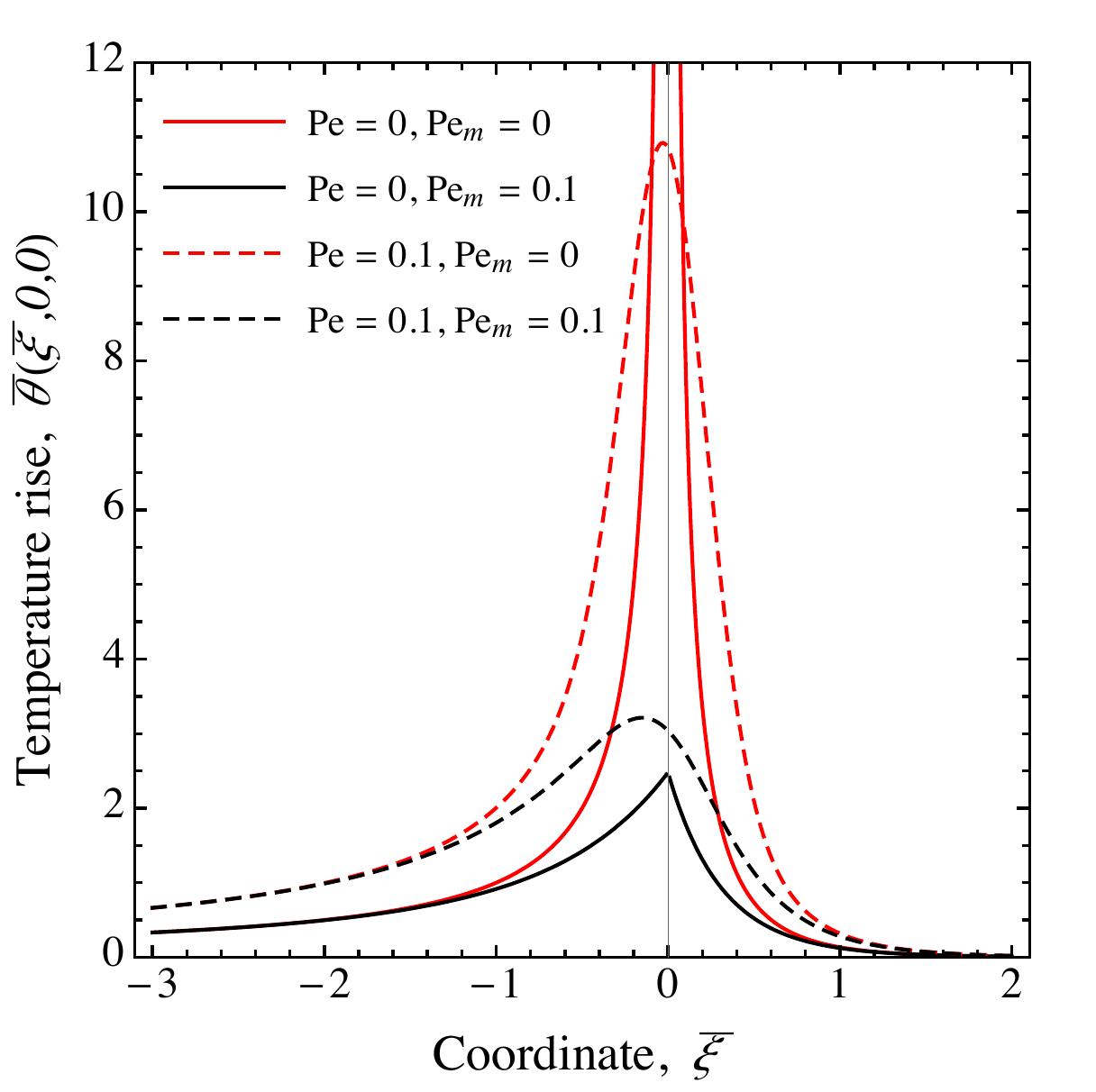}
\caption{Profiles of the temperature rise along the movement direction of the Gaussian heat source for different values of the standard Peclet number Pe and the micro-scale Peclet number Pe$_m$.}
\label{fig4}
\end{figure}

\begin{figure}[b!]
\textbf a
\includegraphics[width=0.5\linewidth]{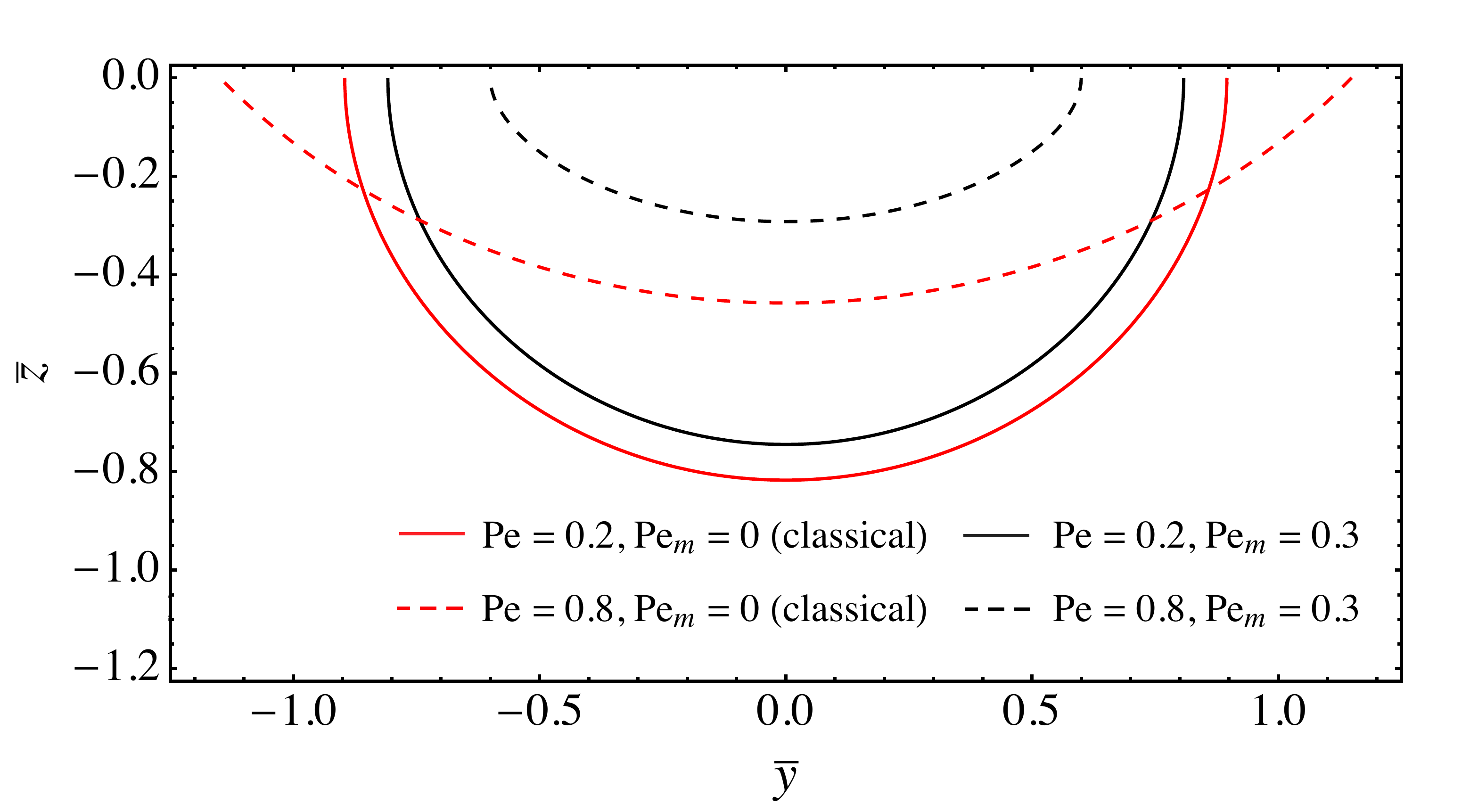}
\textbf b
\includegraphics[width=0.4\linewidth]{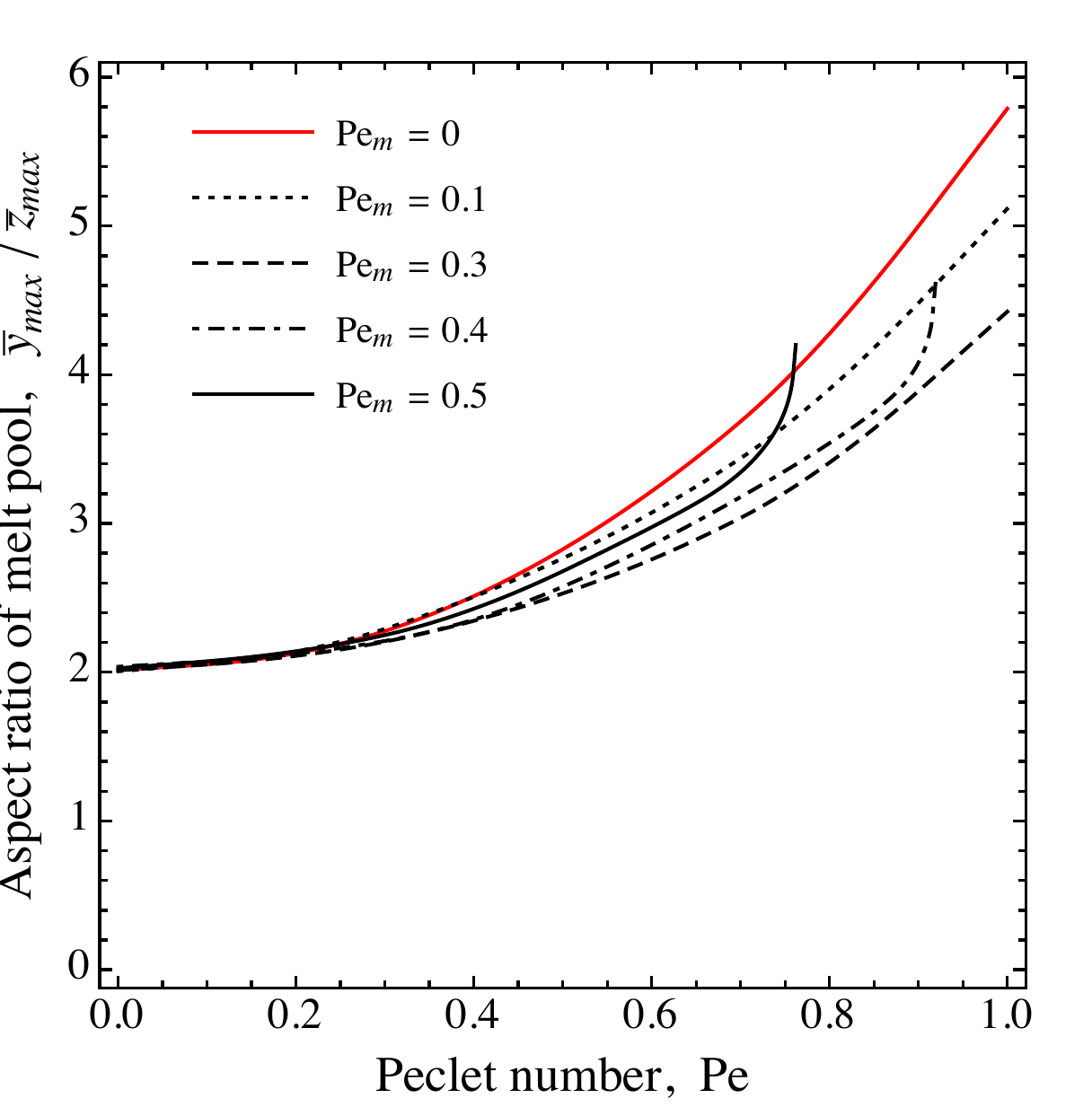}
\caption{Change of melt pool shape (\textbf a) and aspect ratio (\textbf b) in the classical (red lines) and gradient (black lines) solutions for the Gaussian heat source}
\label{fig5}
\end{figure}

The next example of calculations is given for the Gaussian heat source \eqref{conv} in Figs. \ref{fig4}, \ref{fig5}. In Fig. \ref{fig4} we show the influence of the standard and the micro-scale Peclet numbers Pe and Pe$_m$ on the temperature profile along the movement direction. Classical semi-analytical solution with Pe$_m=0$ provides the assessment for the finite level temperature rise, however, the decrease of the temperature with increase of Pe number is not so strong as in the case of Pe$_m$. For the same values of these numbers, classical solution for Gaussian beam (Pe$=0.1$, Pe$_m=0$) predicts the six time higher temperature rise in comparison with those one of gradient solution for the point source (Pe $=0$, Pe$_m=0.1$). Thus, the same values of these non-dimensional parameters provide the changes of different order in the resulting solutions. 

It is interesting to note, that the maximum temperature rise in the gradient Gaussian beam solution (Fig. \ref{fig4}, black dashed line) is higher than those one of the gradient point source solution (Fig. \ref{fig4}, black solid line). This is the consequence of the more intensive heat flux around the center of the laser spot (origin of the coordinates) that is prescribed by the Gaussian distribution \cite{cline1977heat}. Also, the gradient effects do not change the asymptotic far-field behavior of classical solutions nor for the point nor for the Gaussian heat sources.

The influence of the Peclet numbers Pe and Pe$_m$ on the melt pool size and shape under the Gaussian heat source is shown in Fig. \ref{fig5}. Here it is seen the main difference between these non-dimensional numbers. Change of the standard Pe number significantly influenced the aspect ratio (AR) of the melt pool. Its width become larger than the depth (Fig. \ref{fig5}a). In the case of relatively high Pe numbers there arise the aspect ratio up to AR = 6 and higher. In the case of small Pe values the Gaussian beam solution tends to the point source model solution with axial symmetry (AR = 2). Numerical evaluation of the melt pool aspect ratio within the developed solutions was performed according to the following definition:
\begin{equation}
\begin{aligned}
\label{AR} 
AR = \frac{\bar y_{max}}{\bar z_{max}} 
= \left\{\frac{\max |\bar y|}{\max |\bar z|}: \bar\theta(\bar \xi,\bar y,\bar z) = 1\right\}
\end{aligned}
\end{equation}

The non-zero values of Pe$_m$ and corresponding gradient effects  provide the reduction of all dimensions of the melt pool (Fig. \ref{fig5}a). However, the dependence of the melt pool AR is more complex and non-monotonic in gradient solution (Fig. \ref{fig5}b). For a not very large values of Pe$_m$ there arise a slight decrease of the AR value in gradient solution (Fig. \ref{fig5}b, dotted and dashed lines). However, in the case of large values of Pe$_m$ there arise the inverse effect. Moreover, the melting will not arise if Pe$_m$ and Pe values will be too large (Fig. \ref{fig5}b, solid and dot dashed black lines). According to the definitions of these number, this last case corresponds to the situation when the velocity of the heat source $v$ is large, or the material thermal diffusivity $\kappa$ is small, or the characteristic radius of Gaussian beam $a$ is large (the beam is defocused and the heat is distributed over the large area) or the material length scale parameter $l$ is large. In these cases, the melt pool dimensions become small, however its AR may become even larger the in the classical solution.

\begin{figure}[b!]
\centering
\includegraphics[width=0.6\linewidth]{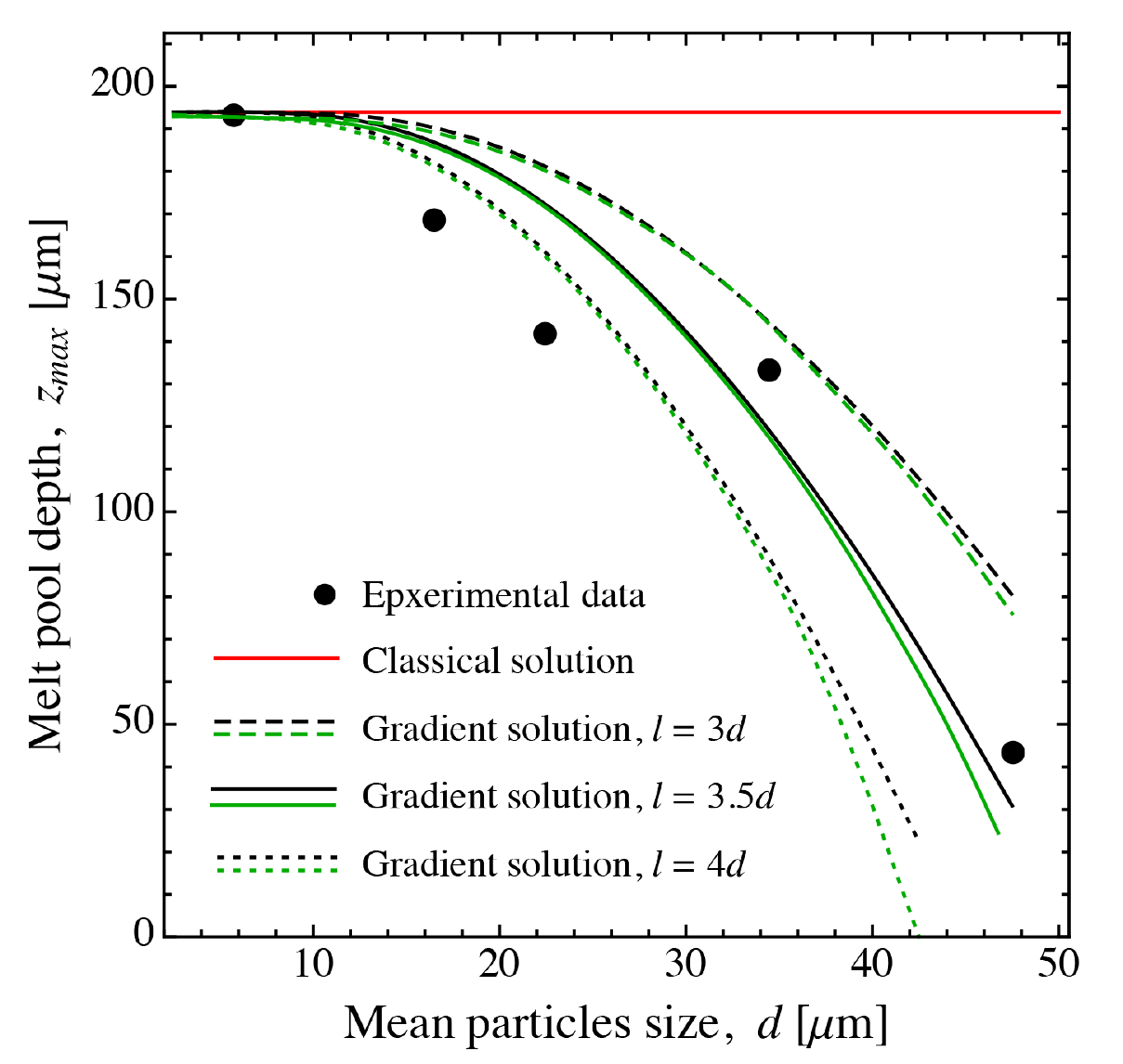}
\caption{Dependence of the melt pool depth on the mean particles size for the selective laser melting of tungsten powders. Red line -- classical models for the point and Gaussian heat source. Black lines -- gradient model of point heat source. Green lines -- gradient model of Gaussian heat source. Experimental data is taken from Ref. \cite{zhang2019influence}.}
\label{fig6}
\end{figure}

Finally, example of comparison of the modeling results with experimental data is shown in Fig. \ref{fig6}. Here we present the dependence of the melt pool depth on the mean particles size for the laser melting of the tungsten powders. Experimental data (black dots in Fig. \ref{fig6}) is taken from Ref. \cite{zhang2019influence}. It is seen, that in the experiments, there arise the decrease of the melt pool depth for the powders with larger particles. Note, that classical models of heat sources cannot be directly used for the description of such experiments due to the lack of the appropriate length-scale parameters. To describe such effects within classical models one should introduce some additional relations for the dependence of the material properties on the powder particles size. Namely, in Ref. \cite{zhang2019influence} it was found that the absorptivity of the tungsten powder may variates in the range of $\lambda=0.5...0.6$ for the used mean sizes of the powder particles $d = 4...50\,\mu$m. However, such variation of absorptivity cannot be used to explain the observed decrease of the melt pool size in about 4 times. In Ref. \cite{zhang2019influence} it was suggested that this effect can be explained by the additional influence of the inhomogeneous irradiation and the peculiarities of the Marangoni flow in the powder beds with large particles. By the other researchers it was also noted that the change of the powder particle size in the laser melting processes may affects the number of competing factors, such that the amount of the contact thermal conductivity inside the powder \cite{tolochko2003mechanisms}, radiation penetration depth and the effective extinction coefficient  \cite{tolochko2003mechanisms, rombouts2005light, gusarov2020radiative, zhang2016selective}, volumetric specific heat \cite{tolochko2003mechanisms}, balling phenomenon \cite{zhang2016selective}, agglomeration, surface state and related kinetics of densification \cite{simchi2004role}. In the present work, we show that the gradient models of moving heat sources can be also used for the continuum-level description of such experimental data by using phenomenological introduction of the relation between the mean particle size and the model's length scale parameter.

The modeling results in Fig. \ref{fig6} were obtained by using the point source model \eqref{GFpn} and the Gaussian source model \eqref{conv}. Firstly, in these solutions we defined the dimensionless coordinates scale ($\bar \xi = v \xi /(2\kappa)$, etc.), and the Peclet numbers  Pe$_m = vl/(2\kappa)$ and Pe $= va/(2\kappa)$. According to the experiments, in these definitions we used the scanning velocity $v=0.2$ m/s and the Gaussian laser beam radius $a=35\, \mu$m. Thermal diffusivity of the tungsten powder $\kappa$ is not available. However, it is known that the thermal diffusivity of the solid tungsten can be approximated by the constant value $\kappa_W = 50$ mm$^2$/s in a wide temperature range \cite{fukuda2018thermal, tanabe2003temperature}. For the powder material the value of thermal diffusivity can be less than those one of the solid material in about 20 times \cite{ahsan2020experimental}. Thus, for the rough assessment, in the calculations we used the value $\kappa = \kappa_W/2 = 25$ mm$^2$/s. This value approximates the change of the powder properties from a very low level at the room temperature up to the relatively high level at the melting point. Based on the made assumptions we defined the dimensionless quantities as follows:
$$
\bar \xi = \frac{v \xi}{2\kappa} = \frac{\xi}{L_0},\quad
\bar R = \frac{v R}{2\kappa} = \frac{R}{L_0},\quad
\bar r = \frac{v R}{2\kappa} = \frac{R}{L_0}
$$
$$
\text{Pe} = \frac{v a}{2\kappa} = \frac{a}{L_0} = 0.14, \quad
\text{Pe}_m = \frac{v l}{2\kappa} = \frac{l}{L_0}
$$
where $L_0 = 2v/\kappa = 250\,\mu$m is the absolute distance that corresponds to the unit value of the dimensionless coordinates.

Then, we assumed that the length scale parameter $l$ can be related to the mean size of the powder particles $d$ that is known from the experiments\cite{zhang2019influence}. We  suppose the linear relation $l=kd$ ($k$ -- proportionality coefficient), such that the micro-scale Peclet number was finally defined as:
$$
\text{Pe}_m = k \frac{d}{L_0}
$$.
The last unknown parameter of the models \eqref{GFpn}, \eqref{conv} is the operating parameter $n$. For the explicit definition of this parameter we need an additional information about the temperature dependent density and heat capacity of the powder. Due to the absence of this data, we identified this parameter $n$ by fitting the classical models to the experiments with the smallest mean powder size $d = 5.7\,\mu$m. We found that point source solution predicts the experimental melt pool depth $z_{max} = 193\,\mu$m when $n=1.4$, while for the classical Gaussian source we found that this parameter should equals $n=0.72$. Then, we used these operating parameters in the corresponding gradient models of moving heat sources to obtain the predictions for the melt pool depth in the case of different mean size of powder particles. 

Note, that the found values of the operating parameters have a typical order. To show this, for example, we may found the approximate theoretical value of the operating parameter by its standard definition as follows
\begin{equation}
\label{oper} 
n_{theor} = \frac{\lambda P v }{4\pi \kappa^2 \rho c(T_m-T_i)} \approx 2.8
\end{equation}
where we use the powder absorptivity $\lambda=0.55$ (mean value that was determined in Ref.\cite{zhang2019influence}), laser power $P=350$ W (used in the experiments\cite{zhang2019influence}), $\kappa = 25$ mm$^2$/s (defined above), $\rho=\rho_w/2 = 9625$ kg/$m^3$ (twice lighter than the solid tungsten), heat capacity $c=c_w \approx 266$ J/(kg K) (twice higher that of the solid tungsten\cite{zhao2016thermal}), tungsten melting point $T_m=3422$ C$^o$ and initial temperature $T_i=20$ C$^o$. 

This assessment \eqref{oper} is given here just for comparison with identified values of $n$ and we did not use theoretical value $n_{theor}=2.8$ in the calculations because this leads to the significant overestimations of the melt pool depth.

Therefore, based on the all assumptions discussed above, in the calculations we used the sets of parameters that are listed in Table \ref{tab1}. We used the experimental values of the mean particles size $d =  \{5.7, 16.52, 22.47, 37.26,$ $ 47.63\}\, \mu$m from Ref.\cite{zhang2019influence} to define the length scale parameter $l=kd$ with three values of proportionality coefficients $k=3, 3.5$ and 4. 

\begin{table}
\caption{Parameters of the models used in the calculations for the selective laser melting experiments with tungsten powders}
\label{tab1}       
\centering
\small
\begin{tabular}{llllllll}
\hline\noalign{\smallskip}
Parameter & \,\quad\,&Dimensions &\quad\,& Value \\
\noalign{\smallskip}\hline\noalign{\smallskip}
Scanning speed, $v$         && m/s  && 0.2         \\
Thermal diffusivity, $\kappa$\quad          & & m$^2$/s  && $25\cdot10^{-6}$      \\
Coordinates scale, $L_0 = 2\kappa/v$\quad   & & $\mu$m  && 250      \\
Pe number       && -  && 0.14$^*$     \\  
Pe$_m$ number && -  && $k d /L_0$      \\ 
Operating parameter, $n$ && - && 1.4 (0.7$^*$) \\
Critical temperature rise, $T_m-T_i$ && $^o$C && 3400\\
\noalign{\smallskip}\hline\\
\end{tabular}\\
$^*${parameters for the Gaussian source model}
\end{table}

Estimated predictions for the melt pool depth are shown in Fig. \ref{fig6} by black lines (gradient point source model) and green lines (gradient Gaussian source model). It is seen, that these models allow to predict the experimentally observed effects of decrease of the melt pool depth with increase of the mean particles size. The most close predictions are obtained if the proportionality coefficient equals to $k=3.5$. The difference between the Gaussian and the point source models is negligible in the present case (since the Peclet number is not very large). Classical predictions are shown by the horizontal red line in Fig. \ref{fig6}. Classical models do not contain additional length scale parameters and cannot predict such type of the size effects without introduction of additional relations for the dependence of the materials properties on the mean particles size.

\section{Conclusion}
\label{con}

In this paper we propose a new simplified phenomenological gradient theory of heat transfer for simulations of the laser powder bed fusion processes. We show, that this theory allow to obtain a generalized solutions for the classical moving source problems that take into account the effects of the material internal characteristic length scale. For the powder bed fusion we propose to related this internal length scale with the mean particle size of the powder. Known experimental data allowed us to validate this assumption and to found that the length scale parameter of the model may have the order of 3.5 of mean particles diameter. 

All gradient effects are incorporated in the derived solutions through the new kind of the non-dimensional group of parameters that we defined here as the micro-scale Peclet number. In the case of small value of the this parameter we obtain the classical solutions, while its large values correspond to the case of the strong influence of the gradient effects and related decrease of the temperature field. Namely, the point and the line source solutions, as well as the Green functions of the presented theory does not contain singularities that can be useful for the practical applications and for the development of more complex numerical/analytical simulation methods.

In general, it seems, that there may exist other variants of the simplified theories that allow to obtain a closed form representation of general solution similar to \eqref{GS}, \eqref{GSo}. Development of such more general theories is the subject for the authors future work as well as the further efforts for the experimental identification of the length scale parameters for different powder materials. \\

%\textbf{Declarations of interest:} The authors declare that they have no conflict of interest.

%\section*{References}
\bibliographystyle{elsarticle-num}
\bibliography{refs.bib}

\end{document}